\documentclass[a4paper,11pt]{article}
\pdfoutput=1 

\usepackage{jcappub} 

\usepackage[dvipsnames]{xcolor}
\usepackage[normalem]{ulem}
\usepackage[utf8]{inputenc}
\usepackage[english]{babel}
\usepackage{microtype}
\usepackage{hyperref}
\hypersetup{
    colorlinks=true,
    pdfborder = {0 0 0.5 [3 3]},
    anchorcolor=black,
    citecolor=blue,
    linktoc=all,
    linktocpage=true,
    linkcolor=red,
    urlcolor=blue
}
\usepackage[capitalize]{cleveref}

\usepackage{siunitx}
\usepackage{amsmath}
\usepackage{graphicx}
\usepackage{mathtools}
\usepackage{booktabs}
\usepackage{bm}

\newcommand{\ud}{\mathrm{d}}
\newcommand{\Planck}{\textit{Planck}}

\newcommand{\taureio}{\tau_{\mathrm{reio}}}
\newcommand{\Astau}{A_s e^{-2 \taureio}}
\newcommand{\fs}{\mathrm{fs}}
\newcommand{\fld}{\mathrm{fld}}
\newcommand{\Nfs}{N_\fs}
\newcommand{\Nfld}{N_\fld}
\newcommand{\Nx}{N_{\mathrm{fs/fld}}}
\newcommand{\Ntot}{N_\mathrm{tot}}
\newcommand{\ffs}{f_\fs}
\newcommand{\rdrag}{r_\mathrm{d}}
\newcommand{\zeq}{z_\mathrm{eq}}
\newcommand{\aeq}{a_\mathrm{eq}}

\newcommand{\Mpc}{\mathrm{Mpc}}
\newcommand{\eV}{\mathrm{eV}}

\newcommand{\YHe}{Y_\mathrm{He}}
\newcommand{\LambdaCDM}{\Lambda\mathrm{CDM}}
\newcommand{\montepython}{\texttt{MontePython}}
\newcommand{\cobaya}{\texttt{cobaya}}
\newcommand{\class}{\texttt{CLASS}}
\newcommand{\fsky}{f_\mathrm{sky}}

\title{Abundance and properties of dark radiation from the cosmic microwave background}

\author[a,1]{Murali M. Saravanan,\note{Corresponding author.}} 
\author[b, c]{Thejs Brinckmann,}
\author[a]{Marilena Loverde,}
\author[a, d]{Zachary J. Weiner}

\affiliation[a]{Department of Physics, University of Washington, Seattle, WA, USA}
\affiliation[b]{Dipartimento di Fisica e Scienze della Terra, Universit\`a degli Studi di Ferrara, Ferrara, Italy}
\affiliation[c]{Instituto Nazionale di Fisica Nucleare (INFN), Sezione di Ferrara, Ferrara, Italy}
\affiliation[d]{Perimeter Institute for Theoretical Physics,  Waterloo, Ontario, Canada}

\emailAdd{msarav@uw.edu}
\emailAdd{thejs.brinckmann@gmail.com}
\emailAdd{mloverde@uw.edu}
\emailAdd{zweiner@perimeterinstitute.ca}

\abstract{
    We study the cosmological signatures of new light relics that are collisionless like standard
    neutrinos or are strongly interacting.
    We provide a simple and succinct rephrasing of their physical effects in the cosmic microwave
    background, as well as the resulting parameter degeneracies with other cosmological parameters,
    in terms of the total radiation abundance and the fraction thereof that freely streams.
    In these more general terms, interacting and noninteracting light relics are differentiated by
    their respective decrease and increase of the free-streaming fraction, and, moreover, the
    scale-dependent interplay thereof with a common, correlated reduction of the fraction of matter
    in baryons.
    We then derive updated constraints on various dark-radiation scenarios with the latest
    cosmological observations, employing this language to identify the physical origin of the impact
    of each dataset.
    The ``PR4'' reanalyses of \textit{Planck} CMB data prefer a larger primordial helium yield and
    therefore also slightly more radiation than the 2018 analysis; we investigate the differences
    between the two releases that drives these shifts.
    Smaller free-streaming fractions are disfavored by the excess lensing of the CMB measured in
    lensing reconstruction data from \textit{Planck} and the Atacama Cosmology Telescope.
    On the other hand, baryon acoustic oscillation measurements from the Dark Energy Spectroscopic
    Instrument drive marginal detections of new, strongly interacting light relics due to that
    data's preference for lower matter fractions.
    Finally, we forecast measurements from the CMB-S4 experiment.
}

\begin{document}
\maketitle
\flushbottom

\newpage

\section{Introduction}
\label{sec:intro}

Measurements of the cosmic microwave background (CMB) precisely determine cosmological parameters and enable tests of the $\Lambda$--cold-dark-matter ($\LambdaCDM$) model against other extended cosmologies~\cite{Planck:2018vyg}.
In particular, the CMB is a powerful probe of physics beyond the Standard Model (BSM) that feature new light particles. Even light relics that interact too weakly with the Standard Model to be detected in laboratory experiments can still be produced at an appreciable level at the high temperatures of the early Universe, making cosmological probes especially powerful. Neutrinos are one example of such a relic found within the Standard Model (SM) itself. Other examples beyond the SM include axions and axionlike particles, sterile neutrinos, models with nonstandard neutrino interactions, and more general dark sector particles~\cite{Turner:1986tb,Baumann:2016wac,Millea:2015qra,Abazajian:2001nj,Weinberg:2013kea,Chacko:2015noa,Buen-Abad:2015ova,Berlin:2017ftj,Dvorkin:2022jyg,Bell:2005dr,Friedland:2007vv,Cyr-Racine:2013jua,Oldengott:2014qra,Kreisch:2019yzn,Wilkinson:2014ksa,Buen-Abad:2017gxg,Archidiacono:2013dua,Archidiacono:2016kkh}.
Since the CMB is a precision probe of the Universe at the end of the radiation-dominated epoch, it is sensitive to this ``dark'' radiation. Even if these light relics have no nongravitational interactions with the SM at times close to recombination, they still leave signatures in the CMB through gravity alone~\cite{Bashinsky:2003tk,Baumann:2015rya}.

The light relics predicted in BSM models need not free stream like SM neutrinos do after the weak interactions decouple.
In some cases, new particles maintain sufficient self-interaction strength within their own sector that they can be treated as tightly coupled; we refer to this as fluidlike radiation (see, e.g., Refs.~\cite{Jeong:2013eza,Buen-Abad:2015ova,Chacko:2015noa}).
Fortunately, the CMB can distinguish between free-streaming and fluidlike radiation and can therefore shed light on broad classes of BSM models while remaining agnostic to their microphysical details. This potential motivates an ongoing effort to study phenomenological parametrizations of dark radiation sectors with current CMB data~\cite{Forastieri:2015paa,Brust:2017nmv,Choi:2018gho,Kreisch:2019yzn,Forastieri:2019cuf,Das:2020xke,RoyChoudhury:2020dmd,Brinckmann:2020bcn,Blinov:2020hmc,RoyChoudhury:2022rva,Brinckmann:2022ajr,Taule:2022jrz,Das:2023npl,Venzor:2023aka,Camarena:2024daj,Poudou:2025qcx,ACT:2025tim}. Searches for additional light degrees of freedom remain an important science driver for future CMB experiments, such as Simons Observatory~\cite{SimonsObservatory:2018koc} and CMB-S4~\cite{CMB-S4:2016ple}.

A clear theoretical understanding of the differences between the fluidlike and free-streaming radiation is required to maximize the discovery potential for future surveys. Building on prior work, we outline how parameter degeneracies differ between models with extra free-streaming radiation versus extra fluidlike radiation.
In \cref{sec:background_cosmo}, we put these scenarios on a common, more transparent footing by
considering general scenarios parametrized by the total radiation content and its partitioning into
free-streaming and fluidlike forms.
We detail two partial degeneracy directions deriving from the distinctive physical effects each parameter controls: a ``tilt degeneracy'' associated with the impact of radiation density on the background cosmology and a ``shift degeneracy'' arising from the impact of free-streaming perturbations on the photon-baryon plasma.
In \cref{sec:constraints} we then provide updated \Planck{} constraints on models with new light
relics using the latest PR4 dataset~\cite{Tristram:2023haj,Carron:2022eyg}, CMB lensing data from the
Atacama Cosmology Telescope~\cite{ACT:2023dou, ACT:2023kun} (ACT), and baryon acoustic oscillation
(BAO) data from the Dark Energy Spectroscopic
Instrument~\cite{DESI:2024lzq,DESI:2024mwx,DESI:2024uvr} (DESI).
Throughout, we apply the language developed in \cref{sec:background_cosmo} to explain the physical
origin of these results, highlighting cases in which particular datasets have a differing impact on
parameter inference for fluidlike and free-streaming light relics.
Lastly, we forecast constraints achieved by the current planned configuration for CMB-S4~\cite{CMB-S4:2016ple,Raghunathan:2023yfe}.
We summarize and conclude in \cref{sec:conclusion}.

\section{Dark radiation and degeneracies in the CMB}
\label{sec:background_cosmo}

The energy density in radiation determines the background expansion rate in the early Universe, controlling physical scales like the sound horizon and the diffusion scale~\cite{Hou:2011ec,Hu:1995en}.
However, the CMB is sensitive not only to the total amount of radiation but also to its properties via the dynamics of spatial perturbations. Perturbations of these additional light relics and their impact on the CMB power spectra are well understood~\cite{Bashinsky:2003tk,Baumann:2015rya,Choi:2018gho}. Broadly speaking, the effects of perturbations are twofold: they impact the amplitude of the CMB power spectrum and shift the angular locations of its extrema.
The phase shift in particular has been studied in detail and measured in data~\cite{Follin:2015hya,Pan:2016zla, Montefalcone:2025unv}.
The magnitude of these effects can be calculated analytically by expanding the Boltzmann equations in the fraction of the total radiation density ($\rho_r$) that freely streams ($\rho_{\mathrm{fs}}$), defined as $\ffs \equiv \rho_{\mathrm{fs}}/\rho_r$.
In this section we review the separate impacts of the total density (\cref{sec:vary_omega_r}) and the free-streaming fraction (\cref{sec:vary-ffs}) on the CMB, remaining agnostic to the actual particle content (aside from the photons).
We then apply these results in \cref{sec:fs_vs_fld} to compare and contrast the signatures of new light relics that freely stream like SM neutrinos or are instead fluidlike.
Before proceeding, we establish notation.

Following convention, we parametrize the contributions to the radiation density as
\begin{align}
    \rho_{\mathrm{r}}
    = \rho_\gamma + \rho_\nu + \rho_{\text{BSM}}
    = \rho_\gamma \left[1 + \frac{7}{8}\left(\frac{4}{11}\right)^{4/3} (\Nfs + \Nfld) \right],
\label{eqn:omega_r_def}
\end{align}
where $\rho_\gamma$, $\rho_\nu$ and $\rho_{\text{BSM}}$ are the energy densities in photons, neutrinos and any BSM sectors.
We define the total effective number of relativistic species as
\begin{align}
    \Ntot \equiv \Nfs + \Nfld
\end{align}
where $\Ntot$ counts the total effective number of degrees of freedom in light relics, $\Nfs$ those that are free-streaming, and $\Nfld$ those that are fluidlike.
Note that $\Nfld$ parametrizes the energy density of fluidlike radiation from BSM particles but not
from photons, which are effectively fluidlike before recombination due to their interactions with
electrons.
The above parametrization of the radiation density allows us to write the free-streaming fraction as
\begin{align}
    \ffs
    \equiv \frac{\rho_{\mathrm{fs}}}{\rho_r}
    = \frac{7/8 \cdot \left(4/11\right)^{4/3} \Nfs}{ 1 + 7/8 \cdot \left(4/11\right)^{4/3} \Ntot}.
\end{align}
In the standard $\LambdaCDM$ model, $\ffs \approx \num{0.4087}$ as the SM predicts $\Ntot = \Nfs = 3.044$.
When allowing for dark radiation in addition to the SM neutrinos' $\Nfs = 3.044$, CMB data alone constrain fluidlike radiation to $\Nfld < 0.47$ at the $95$th percentile but places more stringent constraints on additional free-streaming relics of $\Delta \Nfs < 0.37$ (as we derive in \cref{sec:results-dNfs-dNfld}).
Some current CMB lensing and BAO datasets impact these constraints to a different degree for free-streaming and fluidlike radiation, with some combinations hinting at marginal preferences for the latter (see \cref{sec:Nfld-BAO}).
One of the primary goals of this work is to assess whether these results might indicate the presence of fluidlike light relics or merely reflect discrepancies in parameter inference between different datasets.
Moreover, we forecast how future observations will differentiate between free-streaming and fluidlike radiation.

We define $\omega_b$, $\omega_c$, $\omega_\Lambda$, $\omega_\gamma$, and $\omega_\nu$ as the usual present-day abundances of baryons, cold dark matter, dark energy, photons, and neutrinos, where $\omega_i \equiv \rho_{i, 0} / 3 H_{100}^2 M_{\mathrm{pl}}^2$. We assume a flat Universe and parametrize the Hubble constant by
$H_0 = h \cdot 100~\mathrm{Mpc}^{-1} \mathrm{km/s} \equiv h H_{100}$.
Throughout this work, we fix the summed mass of neutrinos to
$M_\nu \equiv \sum_i m_{\nu_i} = 0.06~\eV$, implemented as one massive neutrino species while all other light relics (both the fluidlike and remaining free-streaming ones) are massless.
When varying the total radiation density $\Ntot$, we ignore the SM prediction for the relic neutrino
abundance; when instead studying additional fluidlike ($\Delta \Nfld$) or free-streaming ($\Delta
\Nfs$) light relics, we do fix the density of SM neutrinos to this prediction, treating them as $3.044$
effective degrees of freedom that freely stream.

Throughout this work, we use the ultrarelativistic fluid species (parametrized by \texttt{N\_{ur}}) implemented in the Boltzmann solver \class{} 3.2.3~\cite{Blas:2011rf,Lesgourgues:2011re,Lesgourgues:2011rh} to model fluidlike radiation.
We set the effective sound speed squared and effective viscosity parameter to $c_{\mathrm{eff, ur}}^2 = 1/3$, and $c_{\mathrm{visc, ur}}^2 = 0$, respectively, which is a standard parametrization for a perfect fluid~\cite{Hu:1998kj}. Note that the ultrarelativistic fluid approximation, which is implemented in \class{} to optimize the modeling of free-streaming radiation at late times, assumes that $c_{\mathrm{visc, ur}}^2 = 1/3$. This approximation must therefore be disabled (by setting the parameter \texttt{ur\_fluid\_approximation} to 3) to accurately model a perfect fluid. We model the standard neutrino content with the implementation of non-cold dark matter species (\texttt{N\_{ncdm}}).
In the remainder of this section, we present a mix of analytic and numerical results.
Numerical degeneracy directions are derived using CMB temperature and polarization data from the \Planck{} PR4 analysis.
We describe the datasets, priors, and methods we employ in \cref{sec:constraints}.

\subsection{Varying radiation density at fixed free-streaming fraction}
\label{sec:vary_omega_r}

We first review the extent to which the parameter freedom of standard $\LambdaCDM$ can compensate for the effects of varying the total radiation density on the primary CMB anisotropies.
The photon density $\omega_{\gamma}$ is measured to extremely high precision via direct measurements of the blackbody distribution~\cite{Fixsen:1996nj, Fixsen:2009ug}, leaving $\omega_r$ free to vary via $\Ntot$ [\cref{eqn:omega_r_def}].
In contrast to prior work, we fix $\ffs$ in this analysis (by adjusting both $\Nfs$ and $\Nfld$ simultaneously) to isolate the variation of the total density in radiation from the changes to the dynamics of its perturbations~\cite{Bashinsky:2003tk, Baumann:2015rya}.
We first review the parameter combinations that are best constrained by CMB and therefore should be held fixed as the radiation density varies, resulting in specific degeneracy directions between the radiation density and other cosmological parameters. Differences in CMB anisotropies that persist along these degeneracy directions are attributed to changes in the pressure-supported fraction of matter~\cite{Hou:2011ec, Ge:2022qws}.

The CMB's sensitivity to $\omega_b$ comes from the equilibrium point between gravitational collapse and radiation pressure in the pre-recombination plasma.
The ratio of the heights of odd peaks (modes that have compressed) and even peaks (modes that have rarefacted) thus effectively measures the ratio of energy density in photons and baryons, commonly quantified as $R(a) = 3 \rho_b / 4\rho_\gamma = a 3\omega_b / 4 \omega_\gamma$ where $a$ is the scale factor. Since we fix $\omega_{\gamma}$, we keep $\omega_b$ fixed for now.

The CMB probes the sum of the baryon and dark matter density $\omega_{cb} \equiv \omega_c + \omega_b$ via the scale factor of matter-radiation equality, $\aeq = \omega_r/\omega_{cb}$. Modes of the primordial plasma that enter the horizon during the radiation era are driven by the decay of the gravitational potentials, referred to as the radiation driving effect~\cite{Hu:1995en}.
The onset of potential decay at horizon crossing coincides with the first oscillation of the plasma and therefore boosts the amplitude of these modes.
But modes that enter the horizon when matter dominates more of the energy budget experience less driving. Therefore, the onset of matter domination determines the impact of radiation driving as a function of scale.
This ``radiation driving envelope'' is sensitive to the fraction of density in radiation over a wide range of redshifts and has a distinct impact on CMB anisotropies~\cite{Knox:2019rjx}.
When varying the radiation density, fixing $\aeq$ via a proportional change to $\omega_{cb}$ (by adjusting $\omega_c$, when $\omega_b$ is fixed) thus isolates the other effects of extra radiation from the well-measured radiation driving effect.

The angular positions of the acoustic peaks of the CMB are precisely quantified by the
angular size of the sound horizon,
\begin{align}
    \theta_s \equiv r_{s, \star} / D_{M, \star},
\end{align}
where $r_{s, \star}$ is the comoving sound horizon at recombination and $D_{M, \star}$ is the comoving distance to the surface of last scattering.
Current data constrain $\theta_s$ with subpermille precision~\cite{Planck:2018vyg}.
The comoving sound horizon is given by
\begin{align}
    r_{s, \star}
    = \int_0^{a_\star} \ud a \frac{c_s(a)}{a^2 H}
    = \int_0^{a_\star} \ud a \frac{c / \sqrt{3[1+R(a)]}}{H_{100} \sqrt{\omega_r + a \omega_{cb} +  + a^4 \omega_\Lambda}},
\end{align}
where $c_s(a) = c / \sqrt{3[1+R(a)]}$ is the sound speed of the plasma and $a_\star$ the scale
factor at recombination.
At such high redshifts ($a \leq a_\star \approx 1/1100$), the contribution from dark energy is negligible.
The sound horizon may then be written in terms of $\aeq$ as
\begin{align}
    r_{s, \star}
    = \frac{c}{H_{100} \sqrt{\omega_r}}
        \int_0^{a_\star} \ud a \frac{1 / \sqrt{3[1+R(a)]}}{\sqrt{1 + a/ \aeq}}.
    \label{eqn:rs-star}
\end{align}
In a Universe with zero mean spatial curvature, the comoving distance to the surface of last scattering is
\begin{align}
    D_{M, \star}
    = \int_{a_\star}^1 \ud a \frac{c}{a^2 H}
    = \int_{a_\star}^1 \ud a \frac{c}{H_{100} \sqrt{\omega_r (1 + a/\aeq )+ a^4 \omega_\Lambda}},
    \label{eqn:DM-star}
\end{align}
to which radiation makes only a small contribution.\footnote{
    Note that by substituting $\omega_m = \omega_{cb} = \omega_r / \aeq$ in \cref{eqn:DM-star},
    we've ignored any additional contributions to the matter density at late times from, e.g.,
    massive neutrinos.
    In this work we fix the sum of neutrino masses to the minimum, $0.06~\mathrm{eV}$;
    the corresponding contribution to the late-time matter density of about half a percent is
    safely ignorable for these analytic estimates.
}
In a flat Universe with $\aeq = \omega_r / \omega_{cb}$ and $\omega_b$ (via $R$) held constant, fixing $\theta_s$ sets the dark energy density $\omega_\Lambda$ (or equivalently the Hubble parameter $h$) as a function of $\omega_r$---specifically, requiring that $h \propto \sqrt{\omega_r}$.

The parameters that remain unspecified are those that determine the curvature power spectrum at the end of inflation and reionization at late times.
We take a standard parametrization of the primordial power spectrum,
\begin{align}
    \Delta_\mathcal{R}^2(k) = A_s \left(\frac{k}{k_p}\right)^{n_s + \alpha_s \ln (k / k_p) / 2 - 1},
\end{align}
where $A_s$, $n_s$, and $\alpha_s$ are the amplitude, tilt, and running of the spectrum, and $k_p$ is a pivot scale conventionally fixed to $k_p = 0.05~\Mpc^{-1}$.
Except where explicitly stated otherwise, we fix the running
$\alpha_s$ to zero.

Varying $\omega_r$ adjusts the size of physical scales, but CMB anisotropies measure the power spectrum as a function of angular scale.
Since $\theta_s$ is effectively fixed by current data (compared to the precision on all other cosmological parameters), the CMB more directly measures the power in the mode that crosses the horizon at recombination, whose wave number is $k_{s, \star}$, rather than any fixed scale like $k_p$.
Scaling $A_s$ to hold $A_s (k_{s,\star}/k_p)^{n_s(k)-1}$ constant across cosmologies thus preserves the initial power at any given angular scale.
Reionization also suppresses the observed amplitude of CMB anisotropies as $A_s e^{-2\taureio}$, where $\taureio$ is the optical depth to reionization, which we fix in this discussion as it is independently constrained by CMB polarization on large scales.

\begin{figure}[t]
    \centering
    \includegraphics[width=\textwidth]{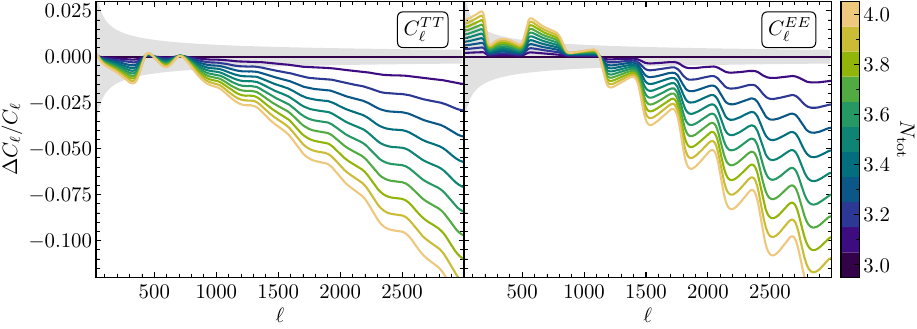}
    \caption{
        Impact of varying the total radiation density $\omega_r$ (i.e., $\Ntot$) at fixed free-streaming fraction $\ffs$ on the unlensed temperature and $E$-mode polarization power spectra.
        Results are displayed as the relative differences from a reference cosmology
        with $\Ntot = 3$.
        Other $\LambdaCDM$ parameters are adjusted to fix $\theta_{s}$, $\aeq$, and the initial power in the mode that crosses the horizon at recombination (with wave number
        $k_{s, \star}$).
        The primordial helium mass fraction is fixed to the BBN prediction in these cosmologies.
        The grey shaded region shows the extent of cosmic variance for observations spanning a sky fraction $\fsky = 0.8$ binned by multipole in intervals $\Delta \ell = 30$.
        }
    \label{fig:vary_omega_r_fix_theta_s}
\end{figure}

To isolate the dependence of the CMB on the total amount of radiation
$\omega_r / \omega_\gamma$ (i.e., $\Ntot$) from changes to its composition, we vary the total
radiation density by adding both fluidlike and free-streaming radiation to fix $\ffs$ to its
$\LambdaCDM$ value of $\num{0.4087}$.
The remaining freedom in $\LambdaCDM$ parameters is used to preserve the aforementioned effects: we fix $100\,\omega_b = \num{2.2218}$, $\zeq = \num{3411}$ (by adjusting $\omega_{c}$), $100\,\theta_{s} = 1.04075$ (by adjusting $\omega_\Lambda$), $\taureio = \num{0.0517}$, and $n_s = \num{0.9635}$.
We scale $A_s$ to fix $\Delta_\mathcal{R}^2(k_{s, \star})$ (taking $A_s = \num{2.0801e-9}$ as a reference value).
\Cref{fig:vary_omega_r_fix_theta_s} shows the effect of increasing $\Ntot$ on both the temperature and polarization power spectra, subject to these choices.
At $\ell \lesssim 1000$, the differences in the temperature spectra are incurred by changes in the pressure-supported matter fraction ($\omega_b/\omega_{cb}$), which we discuss further in \cref{sec:fs_vs_fld}.
While this effect is also relevant on these scales in the polarization spectra, polarization is further enhanced because the width of the visibility function relative to the Hubble rate increases~\cite{Zaldarriaga:1995gi,Baryakhtar:2024rky}, arising from detailed changes to the ionization history as the rate of Thomson scattering per $e$-fold
$\ud \kappa / \ud \ln a \equiv n_e \sigma_T / H$ depends on $\omega_r$.
(Here $n_e$ is the number density of free electrons and $\sigma_T$ the cross-section of Thompson
scattering.)

At smaller scales ($\ell \gtrsim 1000$), on the other hand, both the temperature and polarization
spectra in \cref{fig:vary_omega_r_fix_theta_s} are suppressed by additional radiation.
This scale-dependent suppression is due in part to the broadening of the visibility
function~\cite{Hu:1995em, Zaldarriaga:1995gi, Baryakhtar:2024rky}, but more so due to Silk
damping~\cite{Hou:2011ec}.
Perturbations of the photon distribution on scales smaller than the mean free path of photons during
recombination are exponentially damped by a factor of $e^{-(k/k_D)^2}$, where
\begin{align}
    \label{eqn:k_D}
    k_D^{-2}
    = \int_0^{a_\star}
        \frac{\ud a / a}{(a H)^2}
        \left( \frac{n_e \sigma_T}{H} \right)^{-1}
        \frac{R^2 + 16 (1 + R) / 15}{6 (1 + R)^2}
\end{align}
is the wave number associated with the mean squared diffusion distance at
decoupling~\cite{Hu:1995em, Zaldarriaga:1995gi}.
As established above, the shape of the acoustic peaks motivate fixing $\omega_b$ and $\aeq$,
choices for which $R(a)$ and $H(a) / H(a_\star)$ are fixed as functions of $a / a_\star$.
Then, the variation of $k_D$ with $\omega_r$ (beyond scaling with $a_\star H_\star$) is determined
by the dimensionless Thomson rate $\ud \kappa / \ud \ln a$~\cite{Baryakhtar:2024rky}.
Thomson scattering per $e$-fold decreases in a Universe that expands faster (due to additional
radiation), allowing photons to diffuse further and increasing the damping of small-scale
anisotropies.
As evident in \cref{fig:vary_omega_r_fix_theta_s}, high-resolution CMB observations are quite
sensitive to the angular scale of diffusion, $\theta_D = r_D / D_{M, \star}$, written in terms of
the diffusion length $r_D = 2\pi/k_D$.

To study the parameter dependence of $\ud \kappa / \ud \ln a$, note that most of the photon
diffusion that damps small-scale anisotropies occurs during the last $e$-fold before last
scattering, between helium and hydrogen recombination.
In this period, the number density of free electrons is
$n_e(a) = x_e(a) (1 - \YHe) \rho_b / m_\mathrm{H}$,
where $x_e = n_e / n_\mathrm{H}$ is the hydrogen ionization fraction and
$\YHe \equiv \rho_\mathrm{He} / \rho_b$ the primordial fraction of baryons in helium by mass.
Calculations of Big Bang nucleosynthesis (BBN) in the SM predict the helium yield as a function of
the baryon-to-photon ratio and the radiation density (i.e., $\Ntot$); treating $\YHe$ as a free
parameter is a commonly considered means to compensate for the impact of additional radiation on
diffusion damping~\cite{Hou:2011ec}.
Were the ionization history a fixed function of $a / a_\star$, the diffusion length relative to the
sound horizon would be fixed by taking $1 - \YHe \propto \sqrt{\omega_r}$ (which fixes the Thomson
rate per $e$-fold).
But the ionization history is appreciably modified by variations in $\omega_r$ and $\YHe$, with
nonnegligible impact on the parameter dependence~\cite{Baryakhtar:2024rky}, which we determine
numerically to be
\begin{align}
    r_D / r_s
    &\propto \frac{\omega_r^{0.134}}{\left( 1-\YHe \right)^{0.238}}.
    \label{eqn:rD-rs}
\end{align}
In the SM, the BBN prediction scales approximately as
$1 - \YHe \propto \omega_r^{-0.13}$~\cite{Froustey:2020mcq,Gariazzo:2021iiu}, which compounds
slightly with the scaling of $r_D / r_s$ at fixed $\YHe$ and exacerbates the impact of varying the
radiation density on diffusion damping.

Small-scale damping provides most of the constraining power on $\omega_r$, but freedom in additional beyond-$\LambdaCDM$ parameters, like $\YHe$ as in \cref{eqn:rD-rs}, could conceal this effect.
We next discuss two avenues that compensate for the increase in diffusion damping: adjusting $\LambdaCDM$ parameters beyond the degeneracies discussed in this section (\cref{sec:tilt_degen}) and introducing additional freedom in the primordial power spectrum (\cref{sec:alpha_s}) or the primordial helium yield (\cref{sec:helium}).

\subsubsection{Tilt degeneracy}\label{sec:tilt_degen}

The impact of diffusion damping at large $\ell$ can be partially compensated for by changes to $\LambdaCDM$ parameters, though at the expense of altering the other, aforementioned physical effects (and possibly degrading the fit) at low $\ell$.
For instance, one could modify the scale dependence of the initial conditions and/or the dynamics of the photon-baryon plasma in the radiation era---namely, the scalar tilt $n_s$ and the baryon density $\omega_b$~\cite{Lancaster:2017ksf,Kreisch:2022zxp}.

First, the scale-dependent suppression from diffusion may be compensated for with a bluer initial
power spectrum, i.e., larger $n_s$.
Since the conventional pivot scale corresponds to a multipole of
$\ell_p = k_p / D_{M, \star} \approx 700$, anisotropies at lower multipoles (around the first two
acoustic peaks) are suppressed.
In turn, the first peak may be boosted by increasing $\omega_b$ (i.e., $R$): an increase in
so-called baryon loading shifts the zero point of the acoustic oscillations, enhancing the relative
peak heights.
Increasing the baryon density itself also partly offsets the increased diffusion rate: from
\cref{eqn:k_D}, the diffusion scale increases with $R$, pushing the impact of damping to smaller
scales.

\begin{figure}[!t]
    \centering
    \includegraphics[width=\textwidth]{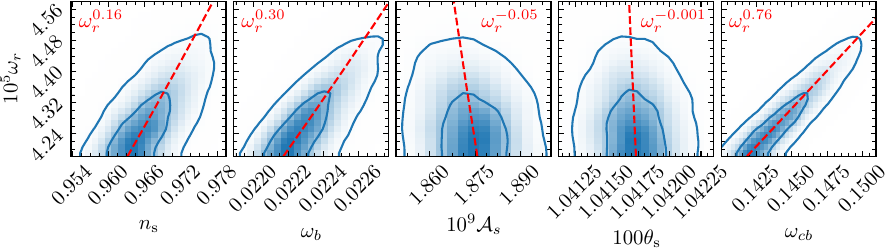}
    \caption{
        Joint posterior distributions between the total radiation density $\omega_r$ and $n_s$, $\omega_b$, $\mathcal{A}_s \equiv \Astau$, $\theta_s$, and $\omega_{cb}$, holding $\ffs$ fixed
        and using the \Planck{} 2020 TT, TE, and EE high-$\ell$ likelihood (CamSpec), the \Planck{}
        2020 low-$\ell$ EE likelihood (\texttt{LoLLiPoP}), and the \Planck{} 2018 low-$\ell$ TT
        likelihood.
        The tilt degeneracy directions, determined via numerical fits, appear in dashed red and are
        labeled on each panel.
        Here $\YHe$ is fixed to the BBN prediction as a function of $\omega_b$ and $\omega_r$.
  }
    \label{fig:tilt_degen_mcmc}
\end{figure}

The competition between these effects---changes to the relative peak heights and the small-scale
suppression due to diffusion---ultimately depends on the relative precision of measurements at large
and small scales for any set of CMB observations, insofar as they determine parameter constraints.
To investigate the relationship between these parameters as constrained by \Planck{} PR4 CMB data, \cref{fig:tilt_degen_mcmc} presents joint posterior distributions (and numerical estimates of degeneracy directions) of $\LambdaCDM$ parameters with $\Ntot$, fixing the free-streaming fraction to $\ffs = 0.4087$.
These results consistently set $\YHe$ to the Big Bang Nucleosynthesis (BBN) prediction, assuming the extra radiation is also present during nucleosynthesis.
(Since we do not change the density in photons, fixing the free-streaming fraction does not allow for $\omega_r$ lower than its standard value, $\omega_\gamma / \ffs$.)

For the posteriors in \cref{fig:tilt_degen_mcmc} deriving from \Planck{} PR4 data, we empirically
find that
\begin{subequations}\label{eqn:tilt-degeneracy}
\begin{align}
    \left. n_s \right\vert_{\ffs}
    \propto \omega_r^{0.16},
    \label{eqn:tilt_degen_n_s}
    \\
    \left. \omega_b \right\vert_{\ffs}
    \propto \omega_r^{0.30}.
    \label{eqn:tilt_degen_omega_b}
\end{align}
\end{subequations}
Since varying the radiation density at fixed $\aeq$ and $\ffs$ does not alter the propagation of acoustic waves in the plasma, $\omega_r$ shows no notable correlation with
$\mathcal{A}_s \equiv \Astau$ and $\theta_s$.
Notably, the matter density exhibits a slightly shallower correlation with $\omega_r$ than would fully fix $\aeq$, due to the competing effect of changing the pressure-supported matter fraction,
$\omega_b / \omega_{cb}$.
However, the relationships in \cref{eqn:tilt-degeneracy} are only slightly modified when extracted
from the subset of the posteriors in \cref{fig:tilt_degen_mcmc} for which $\aeq$ differs by no more
than half a percent from its best-fit value in $\LambdaCDM$.
Since this similarity suggests that the signatures controlled by the pressure-supported matter
fraction are effectively independent, we therefore take $\aeq$ fixed when referring to the tilt
degeneracy below.
We discuss the pressure-supported matter fraction in \cref{sec:helium}, where we promote $\YHe$ to a
free parameter to compensate for the effects of radiation on small-scale damping.
For now, we focus on the correlations of $n_s$ and $\omega_b$ with $\omega_r$.

\begin{figure}[t]
    \centering
    \includegraphics[width=\textwidth]{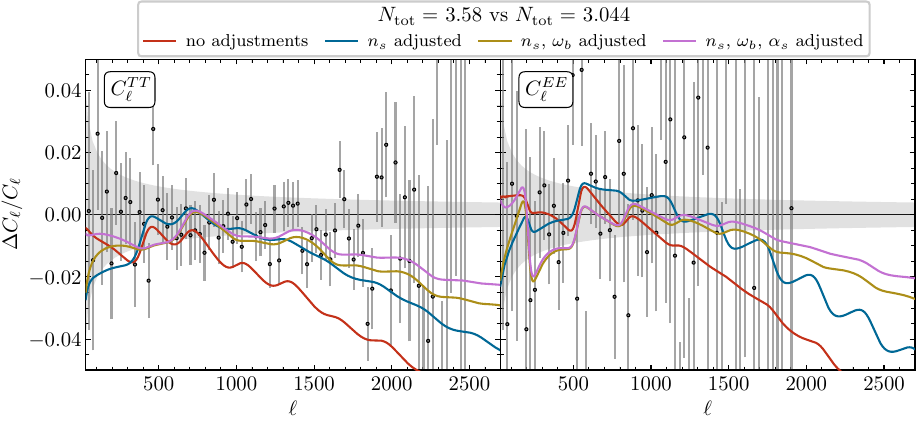}
    \caption{
        Illustration of the tilt degeneracy [\cref{eqn:tilt-degeneracy}], displaying residuals
        of the lensed temperature and $E$-mode polarization power spectra for cosmologies with
        $\Ntot = 3.58$ relative to one with $\Ntot = 3.044$.
        The red, blue, and gold curves respectively adjust no parameters, $n_s$ alone,
        and both $n_s$ and $\omega_b$ according to \cref{eqn:tilt-degeneracy}.
        The purple curve additionally adjusts $\alpha_s$ as in \cref{eqn:alpha_s}.
        Other parameters are specified as in \cref{fig:vary_omega_r_fix_theta_s}.
        The grey shaded region shows the extent of cosmic variance for observations spanning a sky
        fraction $\fsky = 0.8$ binned by multipole in intervals $\Delta \ell = 30$, and the binned
        \Planck{} 2020 spectra and uncertainties are superimposed for the temperature and polarization maps, where all frequencies and their cross spectra have been coadded and the best-fit foreground model subtracted~\cite{Rosenberg:2022sdy}.}
    \label{fig:tilt_degen}
\end{figure}

\Cref{fig:tilt_degen} illustrates the variation in the CMB spectra along the tilt degeneracy.
The relative precision of \Planck{} data at small and large $\ell$ controls the extent to which the tilt degeneracy is viable.\footnote{Throughout this work, we combine the measured $C_\ell$ into bins of width $\Delta \ell = 30$ with centers, values, and uncertainties aggregated with inverse-variance weights. For ease of implementation, we neglect off-diagonal covariance in binning, which underestimates binned uncertainty by no more than $\sim 10\%$. When plotting residuals, we compare to theoretical predictions binned with the same scheme and weights.}
We compare cosmologies with $\Ntot = 3.044$ (the SM prediction) and $3.58$ (the $95$th percentile of
the posterior in \cref{fig:tilt_degen_mcmc}).
Adjusting $n_s$ alone offsets some of the impact on the damping tail but also suppresses the first acoustic peak.
Additionally adjusting $\omega_b$ slightly increases the first peak and also further mitigates the
enhanced damping on small scales.

\subsubsection{Running of the spectral tilt}\label{sec:alpha_s}

Beyond the scalar spectral tilt $n_s$, the running $\alpha_s$ provides additional freedom with which
to compensate for damping induced by additional radiation without altering dynamics.
In addition to the tilt degeneracy [\cref{eqn:tilt-degeneracy}], for the same set of \Planck{} PR4 data we empirically find that the posterior is oriented best along the line
\begin{align}
    \left. \alpha_s \right\vert_{\ffs}
    \propto 3984 \, \omega_r,
    \label{eqn:alpha_s}
   \end{align}
since $\alpha_s$ is constrained close to zero.
The impact of adjusting $\alpha_s$ is also illustrated in \cref{fig:tilt_degen}.
Similar to $n_s$, $\alpha_s$ offsets impacts to the damping tail but, unlike $n_s$, can
simultaneously improve the fit at larger scales.

\subsubsection{Degeneracy with the helium fraction}\label{sec:helium}

\begin{figure}[t]
    \centering
    \includegraphics[width=\textwidth]{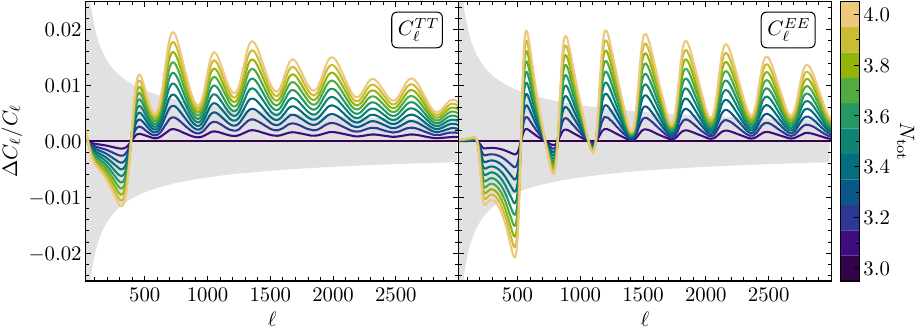}
    \caption{
        Impact of varying the total radiation density $\omega_r$ (i.e., $\Ntot$) on the unlensed
        temperature and $E$-mode polarization power spectra, fixing the free-streaming fraction
        and the angular scale of diffusion $\theta_D = \num{3.23e-3}$
        (by adjusting the Helium mass fraction $\YHe$) as well as $\theta_s$, $\aeq$, and $\Delta_{\mathcal{R}}^2(k_{s, \star})$ as in Fig.~\ref{fig:vary_omega_r_fix_theta_s}.
        The latter controls for the impact of larger radiation densities on small-scale damping,
        leaving the fraction of the matter density that is pressure supported as the most
        significant variable remaining.
        Results are presented as in \cref{fig:vary_omega_r_fix_theta_s}.
    }
    \label{fig:vary_omega_r_fix_theta_s_theta_d}
\end{figure}

As seen in the tilt degeneracy, the standard $\LambdaCDM$ model only has sufficient freedom to
partially compensate for the radiation density's impact on diffusion damping.
\Cref{fig:vary_omega_r_fix_theta_s_theta_d} shows that fixing $\theta_D$ by varying $\YHe$
[according to \cref{eqn:rD-rs}] largely removes the small-scale suppression incurred by larger
radiation densities.
The remaining residuals derive from the change to the fraction of the matter density that is
pressure supported ($\omega_b/\omega_{cb}$)~\cite{Hou:2011ec, Ge:2022qws}, since fixing both $\aeq$
and $R$ with increasing $\omega_r$ requires holding $\omega_b$ constant and increasing $\omega_c$.
Gravitational potentials then decay to a lesser extent in the radiation era, with twofold (and
scale-dependent) effects on acoustic oscillations.
First, the radiation driving effect is diminished for large-scale modes (those that undergo no more
than a single oscillation), reducing the amplitude of photon perturbations around the first
acoustic peak ($\ell \lesssim 400$ in \cref{fig:vary_omega_r_fix_theta_s_theta_d}).
However, smaller-scale modes that oscillate multiple times before last scattering do so in deeper
potential wells, in which they compress to higher densities.
CMB anisotropies are thus enhanced on the smaller angular scales that are primarily sourced by these
modes, as evident in \cref{fig:vary_omega_r_fix_theta_s_theta_d}.
We discuss this effect in connection to the dynamics of radiation perturbations in
\cref{sec:f_fs_vs_pressure_supp}.

\subsection{Varying free-streaming fraction at fixed radiation density}\label{sec:vary-ffs}

\Cref{sec:vary_omega_r} considers the degeneracies of the total radiation density $\omega_r$ with
other cosmological parameters while fixing its composition---that is, the fraction thereof that
freely streams ($\ffs$).
These degeneracies depend largely on the background cosmology rather than the dynamics of its
perturbations.
We now review the impact of the composition of radiation at fixed density ($\Ntot$), following
Refs.~\cite{Bashinsky:2003tk,Baumann:2015rya}.

The effect of free-streaming radiation can be understood analytically in the radiation-dominated
epoch by expanding in small $\ffs$ to study how differences in the gravitational potential sourced
by the anisotropic stress of free-streaming radiation modify the monopole of the photon
distribution.
This correction to the Sachs-Wolfe effect imprints in the amplitude and the location of the extrema of the CMB spectra~\cite{Montefalcone:2025unv}.
The analytic prediction in the radiation-dominated era for the ratio of the amplitudes of the temperature and polarization spectra between first and zeroth order is~\cite{Bashinsky:2003tk, Baumann:2015rya}
\begin{align}
    \label{eqn:amp_shift}
    \frac{C_\ell^{(1)}}{C_\ell^{(0)}}
    = (1 - 0.268 \ffs)^2,
\end{align}
and the acoustic peaks shift in multipole by
\begin{align}
\label{eqn:phase_shift}
    \delta \ell
    \approx - 0.19 \ffs \Delta \ell,
\end{align}
where $\Delta \ell \approx \pi / \theta_s \approx 300$ is the average separation between the peaks~\cite{Pan:2016zla}.

\begin{figure}[t]
    \centering
    \includegraphics[width=\textwidth]{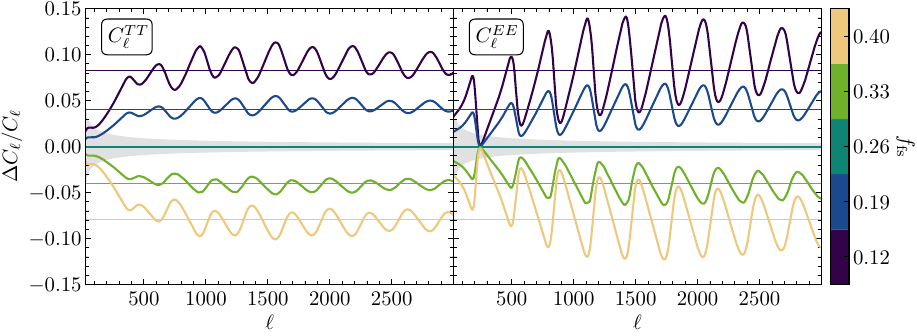}
    \caption{
        Impact of varying the free-streaming fraction $\ffs$ on the unlensed CMB temperature and
        polarization power spectra at fixed $\Ntot = 3$, changing no other cosmological parameters
        (i.e., holding the background cosmology fixed).
        Residuals are computed relative to a cosmology with $\ffs = 0.26$ for illustration.
        Results are depicted as in \cref{fig:vary_omega_r_fix_theta_s}.
        The only physical difference is in the dynamics of radiation perturbations; in contrast to,
        e.g., \cref{fig:vary_omega_r_fix_theta_s_theta_d}, the fraction of pressure-supported matter
        does not vary.
        The colored horizontal lines correspond to the analytic prediction for the amplitude shift with $\ffs$ [\cref{eqn:amp_shift}].
    }
    \label{fig:vary_f_fs}
\end{figure}

\Cref{fig:vary_f_fs} illustrates these physical effects for cosmologies with $\ffs$ varied at fixed
$\Ntot = 3$ to remove the effects of extra radiation density on the background cosmology
(\cref{sec:vary_omega_r}).
The offset in amplitude matches \cref{eqn:amp_shift} well at $\ell \gtrsim 700$, corresponding to
modes for which the approximation of radiation domination is accurate.
The oscillations of the residuals about \cref{eqn:amp_shift} increase in amplitude for greater
changes in $\ffs$ due to the increased phase shift [\cref{eqn:phase_shift}].
The effect of changing $\ffs$ gradually diminishes in the matter era, as radiation's relative
contribution to the Einstein equations decreases.
The residuals in \cref{fig:vary_f_fs} therefore decrease at lower multipoles---those dominated by
scales that only become dynamical around or after matter-radiation equality.
We discuss the scale dependence of these effects further in \cref{sec:f_fs_vs_pressure_supp}.

\subsubsection{Shift degeneracy}
\label{sec:shift_degen}

\begin{figure}[!t]
    \centering
    \includegraphics[width=\textwidth]{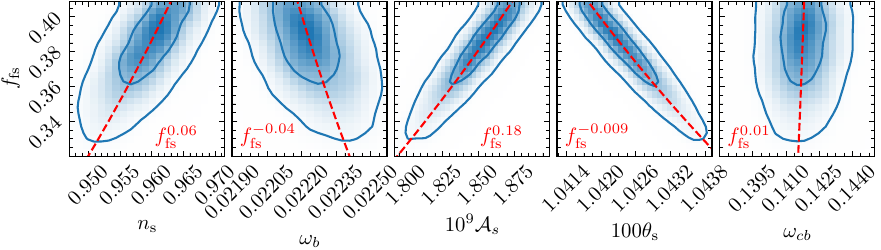}
    \caption{
        Joint posterior distribution between the free-streaming fraction $\ffs$ and $n_s$, $\omega_b$, $\mathcal{A}_s$, $\theta_s$, and $\omega_{cb}$.
        Results fix $\Ntot = 3.044$ and use the \Planck{} 2020 TT, TE, and EE high-$\ell$ likelihood
        (CamSpec), the \Planck{} 2020 low-$\ell$ EE likelihood (\texttt{LoLLiPoP}), and the
        \Planck{} 2018 low-$\ell$ TT likelihood.
        The tilt degeneracy directions, determined via numerical fits, appear in dashed red and are
        labeled on each panel.
        Here $\YHe$ is fixed to the BBN prediction as a function of $\omega_b$.
        }
    \label{fig:shift_degen_mcmc}
\end{figure}

\begin{figure}[t]
    \centering
    \includegraphics[width=\textwidth]{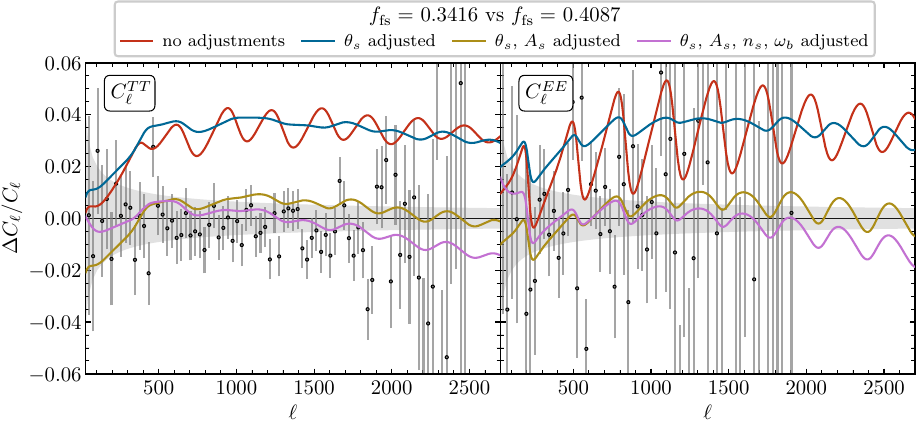}
    \caption{
        Illustration of the shift degeneracy [\cref{eqn:shift-degeneracy}], displaying residuals
        of the lensed temperature and $E$-mode polarization power spectra for cosmologies with
        $\ffs = 0.3416$ relative to one with the SM predicted value of $\ffs = 0.4087$.
        The red, blue, and gold curves respectively adjust no parameters, $\theta_s$ alone,
        and both $\theta_s$ and $A_s$ according to \cref{eqn:shift_degen_A_s,eqn:shift_degen_theta_s}.
        The purple curve additionally adjusts $n_s$ and $\omega_b$ as in \cref{eqn:shift_degen_n_s,eqn:shift_degen_omega_b}.
        All other parameters are fixed.
        Results are otherwise presented as in \cref{fig:tilt_degen}.
    }
    \label{fig:shift_degen}
\end{figure}

Since the free-streaming fraction impacts both the amplitude and the location of the peaks of the spectra, it should be at least partially degenerate with $\mathcal{A}_s \equiv A_s e^{-2\taureio}$, which controls the overall amplitude of CMB anisotropies, and $\theta_s$, which is determined by the location of the peaks~\cite{Blinov:2020hmc,Baumann:2015rya,Pan:2016zla}.
To investigate the relationship between $\ffs$ and other $\LambdaCDM$ parameters,
\cref{fig:shift_degen_mcmc} presents study posteriors deriving from \Planck{} PR4 CMB data in models
where the free-streaming fraction is allowed to vary but the total amount of radiation is fixed to
$\Ntot = 3.044$.
Since we do not alter the photon density and $\Ntot$ is fixed, values of $\ffs$ above the SM prediction of $\ffs = 0.4087$ are disallowed.
As anticipated~\cite{Bashinsky:2003tk,Baumann:2015rya}, we empirically find a strong degeneracy between $\ffs$, $\mathcal{A}_s$, and
$\theta_s$.
We coin this partial numerical degeneracy a ``shift degeneracy'' to differentiate from the tilt
degeneracy (\cref{sec:tilt_degen}) due to the effects of the radiation density on the
background cosmology.
We also find mild correlations with $\omega_b$ and $n_s$, which can both partially mitigate the effects of varying $\ffs$ on the first peak (see \cref{fig:vary_f_fs}).
Unlike for a varying radiation density, for which a proportional increase in $\omega_{cb}$ fixes $\aeq$, there is little correlation between $\ffs$ and $\omega_{cb}$.
For the posteriors in \cref{fig:shift_degen_mcmc} deriving from \Planck{} PR4 data, we empirically
find that
\begin{subequations}\label{eqn:shift-degeneracy}
\begin{align}
    \left. \mathcal{A}_s \right\vert_{\omega_r}
    &\propto \ffs^{0.18},
    \label{eqn:shift_degen_A_s}
    \\
    \left. \theta_s \right\vert_{\omega_r}
    &\propto \ffs^{-0.0090},
    \label{eqn:shift_degen_theta_s}
    \\
    \left. n_s \right\vert_{\omega_r}
    &\propto \ffs^{0.06},
    \label{eqn:shift_degen_n_s}
    \\
    \left. \omega_b \right\vert_{\omega_r}
    &\propto \ffs^{-0.04}
    \label{eqn:shift_degen_omega_b}
    .
\end{align}
\end{subequations}
Note that while the correlation in \cref{eqn:shift_degen_theta_s} appears weak, the exquisite precision of the measurement of $\theta_s$ makes it relevant.
To illustrate this shift degeneracy, \cref{fig:shift_degen} shows the impact of each of $\theta_s$,
$\mathcal{A}_s$, $n_s$ and $\omega_b$ on the CMB power spectra for $\ffs = 0.3416$ (the 5th percentile of the
posterior in \cref{fig:shift_degen_mcmc}) compared to a baseline cosmology with the SM value of $\ffs = 0.4087$.
Adjusting $\theta_s$ compensates for the shift of the peaks, especially in the range $500<\ell<1500$, reducing the oscillations of the residuals in both temperature and polarization.
Shifting $A_s$ accounts for the change in amplitude as predicted in \cref{eqn:amp_shift}.
Finally, adjusting $n_s$ tilts the spectra to better absorb the scale dependent amplitude change in the first peak while adjusting $\omega_b$ further decreases the amplitude in the first peak.
We stress again that this partial degeneracy direction is a function of the precision of \Planck{} observations, which are displayed in \cref{fig:shift_degen}.

\subsection{Free-streaming versus fluidlike radiation}
\label{sec:fs_vs_fld}

We have thus far studied parameter degeneracies when varying the total radiation density at fixed free-streaming fraction and vice versa.
Although $\omega_r$ and $\ffs$ most cleanly encode the physical effects on cosmological observables, contributions from BSM physics are more conveniently parametrized through $\Delta \Nfs$ and $\Delta \Nfld$, insofar as new degrees of freedom either freely stream or are fluidlike (and therefore modify $\omega_r$ and $\ffs$ simultaneously).
(We do not consider the possibility of sectors whose interactions decouple or recouple on
cosmologically relevant timescales~\cite{Choi:2018gho, Park:2019ibn, Kreisch:2019yzn,
Loverde:2022wih, Brinckmann:2022ajr, Kreisch:2022zxp}.)

While equal increases in $\Nfs$ and $\Nfld$ yield the same total radiation density and therefore
increase damping (\cref{fig:vary_omega_r_fix_theta_s}) and decrease the fraction of matter that is
pressure supported (\cref{fig:vary_omega_r_fix_theta_s_theta_d}) to the same extent, they change the
free-streaming fraction by differing amounts and with opposite sign.
To leading order,
\begin{align}
    \Delta \ffs
    \approx \left( 1 - \ffs \right) \frac{\Delta \rho_{\fs}}{\rho_{r}}
        - \ffs \frac{\Delta \rho_{\fld}}{\rho_{r}}.
    \label{eqn:perturb_ffs}
\end{align}
In the baseline cosmology where $\rho_{\fs} = \rho_\nu$ and $\rho_r = \rho_\gamma + \rho_\nu$, the
free-streaming fraction $\ffs\approx 0.4087$; effects that are controlled by the free-streaming
fraction are then roughly $(1 - \ffs) / \ffs \approx 1.45$ times greater in magnitude for an
increase of $\Delta \Nfs$ compared to an equal increase of $\Delta \Nfld$.
When adjusting, e.g., $\YHe$ in order to compensate for the common effect on small-scale damping and
fixing parameters as in \cref{sec:vary_omega_r}, the ultimate impacts of additional free-streaming
and fluidlike radiation are differentiated by the interplay between their common change to the
pressure-supported matter fraction and their distinct changes to the free-streaming fraction.

\subsubsection{Interplay between free-streaming radiation and pressure supported matter}
\label{sec:f_fs_vs_pressure_supp}

\begin{figure}[t]
    \centering
    \includegraphics[width=\textwidth]{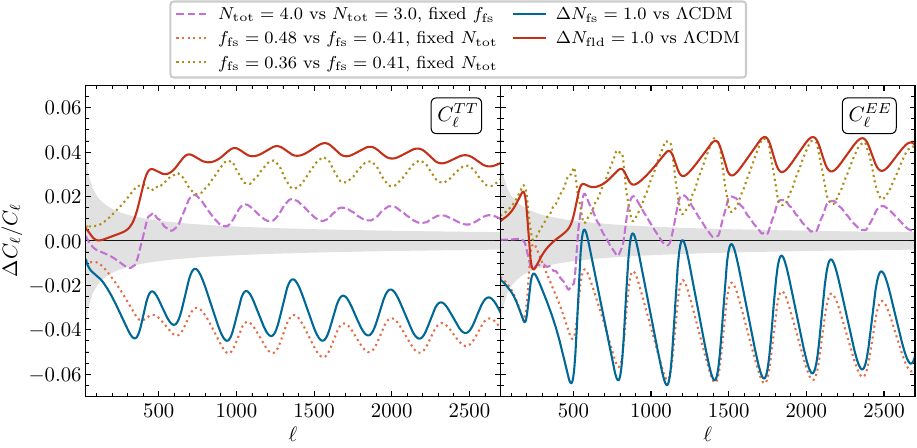}
    \caption{
        Impact of additional free-streaming versus additional fluidlike radiation on the unlensed CMB temperature (left) and polarization (right) power spectra, comparing residuals for cosmologies $\Delta \Nfs = 1$ (blue) and $\Delta \Nfld= 1$ (red) relative to the baseline cosmology.
        In both panels, purple dashed curves increase $\Ntot$ by one but fix $\ffs$; by also fixing
        $\aeq$, $\theta_s$, and $\theta_D$, these results illustrate the isolated effect of changing
        the fraction of pressure-supported matter.
        Dotted curves show the effect of increasing (orange) or decreasing (gold) the
        free-streaming fraction by the same amount incurred by $\Delta \Nx = 1$ while holding
        $\Ntot$ fixed.
        In all cases, $\aeq$, $\theta_s$, $\theta_D$, $\omega_b$, $A_s$, and $n_s$ are fixed
        as in \cref{fig:vary_omega_r_fix_theta_s_theta_d}; results are otherwise presented as in
        \cref{fig:vary_omega_r_fix_theta_s}.
    }
    \label{fig:pressure_supp_vs_f_fs_effects_C_ell}
\end{figure}

To study the impact of correlated changes to the free-streaming fraction and pressure-supported
matter fraction, \cref{fig:pressure_supp_vs_f_fs_effects_C_ell} displays residuals of the
temperature and polarization spectra for cosmologies with $\Delta \Nfs = 1$ and $\Delta \Nfld = 1$
relative to the baseline case with $\Nfs = 3.044$, fixing $\theta_s$, $\omega_b$, $\aeq$, and
$\theta_D$ (by adjusting $\YHe$) as in \cref{sec:helium}.
As discussed in \cref{sec:vary_omega_r}, these choices effectively compensate for all effects
that derive from changes to the expansion history.
\Cref{fig:pressure_supp_vs_f_fs_effects_C_ell} also presents results for commensurate changes to
$\ffs$ at fixed $\Ntot$ and vice versa.\footnote{
    In order to vary $\ffs$ at fixed $\Ntot$ without modifying the photon density, we set
    $\Ntot = 4$ in these cases and adjust other parameters to preserve the $\LambdaCDM$ values of
    $\omega_b$, $\theta_s$, $\theta_D$, and $\aeq$.
}
Note that Ref.~\cite[see Fig. 3]{Ge:2022qws} also compared the signatures of extra fluidlike or
free-streaming radiation to the case with the same radiation density but fixed free-streaming
fraction, interpreting the results relative to an expanded ``mirror-world'' dark sector scenario.
In \cref{fig:pressure_supp_vs_f_fs_effects_C_ell}, we instead consider the complementary case with
fixed radiation density and varying free-streaming fraction in order to explain the contrast in the
combined effects for fluidlike and free-streaming light relics.
We discuss three notable features in \cref{fig:pressure_supp_vs_f_fs_effects_C_ell}:
signatures that are oscillatory in $\ell$ and impact the peak locations, the overall shift in
amplitude at high $\ell$, and the differing effects on the first peak ($\ell \lesssim 400)$.

First, with additional free-streaming radiation the residuals of the temperature spectra
exhibit oscillations in multipole that are comparable in amplitude to their average.
By contrast, when adding fluidlike radiation the residuals' oscillations at $\ell \gtrsim 500$ are
about a tenth (rather than nearly half) of the size of the scale-independent shift.
A similar comparison holds for polarization, but the residuals oscillate with roughly double the
amplitude in all cases.
Decreasing the pressure-supported matter fraction moves the peaks to slightly larger scales by
shifting the zero-point of oscillations in the baryon-photon fluid.
This effect adds constructively with the phase shift induced by increasing the free-streaming
fraction with additional free-streaming radiation, while the two effects are out of phase with
additional fluidlike radiation, exacerbating the oscillations in the former case and diminishing
them in the latter.
The resulting shifts in peak locations due to free-streaming and fluidlike radiation, listed in
\cref{tab:fs_vs_fluid_peak_locs}, therefore differ in magnitude by much more than $1.45$ (and even
in sign for the second and third peak), as predicted if $\omega_b / \omega_{cb}$ were fixed.
\begin{table}[t]
    \centering
    \renewcommand{\arraystretch}{1.2}
    \begin{tabular}{l rrrrrrrrrr}
        \toprule
        & 0 & 1 & 2 &3 & 4 & 5 & 6
        \\
        \midrule
        $\Nfs = 3.044$& 220.7&537.0&815.3&1130.6&1428.4&1740.9&2040.4
        \\
        \midrule
        $\Delta \Nfs = 1.0$& -1.3&-1.7&-2.7&-2.6&-3.0&-2.9&-3.3
        \\
        $\Delta \Nfld = 1.0$ &0.2&-0.1&-0.1&0.3&0.4&0.5&0.7
        \\
        \bottomrule
    \end{tabular}
    \caption{
        Peak locations (in multipole) of the CMB temperature spectrum in $\LambdaCDM$ cosmology
        (top row) and the shift in locations for the $\Delta \Nx =1.0$ models depicted in \cref{fig:pressure_supp_vs_f_fs_effects_C_ell}
        (second and third rows).
    }
    \label{tab:fs_vs_fluid_peak_locs}
\end{table}

Next, in contrast to the expected ratio of $1.45$ from \cref{eqn:perturb_ffs},
\cref{fig:pressure_supp_vs_f_fs_effects_C_ell} displays an overall shift in amplitude at high $\ell$
that is slightly \textit{smaller} in magnitude for free-streaming radiation than that for fluidlike
radiation.
The results in \cref{fig:pressure_supp_vs_f_fs_effects_C_ell} that vary $\ffs$ at fixed $\Ntot$ on
the other hand are indeed consistent with the expectation from \cref{eqn:perturb_ffs}.
This reversal is due again to the common decrease of the fraction of matter that is pressure
supported ($\omega_b / \omega_{cb}$), which \cref{fig:vary_omega_r_fix_theta_s_theta_d} shows boosts
power on small scales by a percent or so (for $\Delta \Ntot = 1$), partly canceling the amplitude
effect under additional free-streaming radiation and compounding with it for fluidlike radiation.

\begin{figure}[t]
    \centering
    \includegraphics[width=\textwidth]{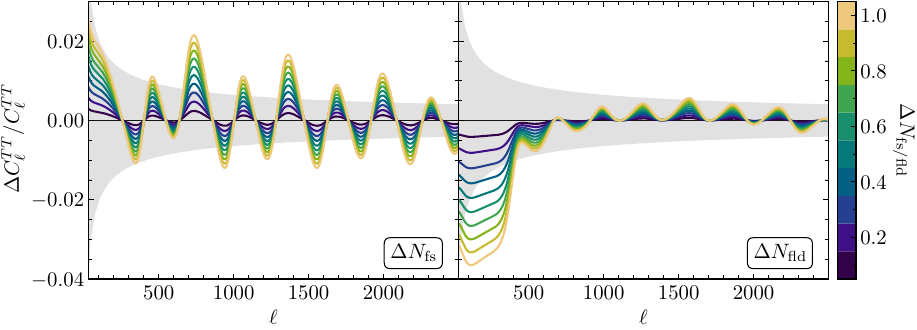}
    \caption{
        Illustration of the partial degeneracy between $\Nx$, $\YHe$, and $A_s$ in the CMB temperature spectra.
        Here $A_s$ is chosen in each case to minimize the mean-squared difference (weighted by
        cosmic variance) of the temperature spectrum compared to the fiducial case with $\Nfs =
        3.044$ for $\ell>1000$.
        Other cosmological parameters are adjusted to fix $\theta_s$, $\theta_D$, $\aeq$, $\omega_b$, $\taureio$, and $n_s$ (see \cref{sec:vary_omega_r}).
        Results are otherwise presented as in \cref{fig:vary_omega_r_fix_theta_s}.
        Although not shown, the polarization spectra exhibit the same key features.
    }
    \label{fig:vary_n_x_fix_theta_s_theta_d_optimized_A_s}
\end{figure}

The interplay of these effects significantly impacts the degeneracy with the amplitude of the
primordial power spectrum, $A_s$.
From \cref{eqn:amp_shift}, to first order in $\ffs$, we expect an overall, fractional change in the
amplitude of the CMB spectra at high multipole of
$\Delta C_\ell(\ffs) / C_\ell(\ffs) = - 0.246 \Delta \ffs / \ffs,$
which in principle could be compensated for by an equal and opposite change $\Delta A_s / A_s$.\footnote{
    This prediction differs from the relationship observed between $\mathcal{A}_s$ and $\ffs$ in the shift degeneracy (\cref{fig:shift_degen_mcmc}) due to the influence of
    low-$\ell$ data, which we discuss next.
    \Cref{fig:vary_f_fs} demonstrates that amplitude differences incurred by changing $\ffs$ only asymptote to the analytic prediction for $\ell \gtrsim 700$.
}
To study the impact of the reduction in $\omega_b / \omega_{cb}$ on degeneracies with the primordial
amplitude, \cref{fig:vary_n_x_fix_theta_s_theta_d_optimized_A_s} varies $\Delta \Nx$ (as usual, at
fixed $\theta_s$, $\aeq$, and $\theta_D$) and adjusts $A_s$ to compensate for the effects at high $\ell$.
We empirically determine the optimal relationship between $A_s$ and $\ffs$ by minimizing the
cosmic-variance-weighted, mean-squared deviation between the temperature spectra of cosmologies of
increasing $\Delta \Nx$ and $\LambdaCDM$ (i.e., $\Nfs = 3.044$) using the Nelder-Mead minimization
algorithm.
The optimized values of $A_s$ approximately lie along
\begin{align}
    \label{eqn:As_vs_eps_X}
    \Delta A_s / A_s =
    \begin{cases}
        0.21 \, \Delta \ffs / \ffs, & \text{for varying $\Delta \Nfs$}
        \\
        0.31 \, \Delta \ffs / \ffs, & \text{for varying $\Delta \Nfld$}
	\end{cases}
\end{align}
which is shallower (steeper) than the degeneracy direction predicted above for free-streaming
(fluidlike) radiation.

As in \cref{fig:pressure_supp_vs_f_fs_effects_C_ell}, the reduction in the pressure-supported matter
fraction when increasing $\Nfs$ amplifies the acoustic oscillations on small scales, partially
compensating for the reduction in amplitude incurred by increasing $\ffs$.
Moreover, for the amplitudes that minimize the deviation at $\ell > 1000$, the residuals at almost all
$\ell$ also oscillate about zero for the case of additional free-streaming radiation.
For fluidlike radiation, on the other hand, the same optimization procedure is ineffective because
of a more severe mismatch between the first acoustic peak and small-scale anisotropies.
The right panel of \cref{fig:vary_n_x_fix_theta_s_theta_d_optimized_A_s} shows that while the optimization does indeed center the residuals on zero at high $\ell$, it does so at the expense of a severely suppressed first peak in the temperature power spectrum.
The same effect is evident in the polarization spectra.

\begin{figure}[t]
    \centering
    \includegraphics[width=\columnwidth]{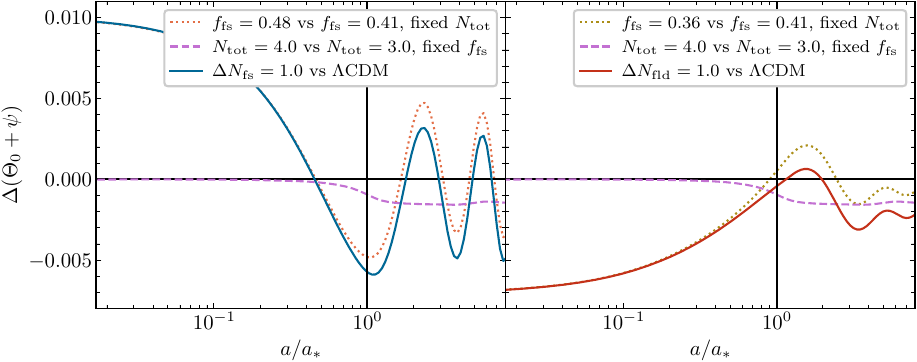}
    \caption{
        Differences in the evolution of the Sachs-Wolfe term $\Theta_0 + \psi$ for a mode that contributes significantly to the first peak of the CMB primary spectra.
        Solid curves compare models with $\Delta \Nfs = 1$ (left, blue) and $\Delta \Nfld = 1$
        (right, red) to $\LambdaCDM$.
        In both panels, purple dashed curves increase $\Ntot$ by one but fix $\ffs$; by also fixing
        $\aeq$, $\theta_s$, and $\theta_D$, these results illustrate the isolated effect of changing
        the fraction of pressure-supported matter.
        Dotted curves show the effect of increasing (left, orange) or decreasing (right, gold) the
        free-streaming fraction by the same amount incurred by $\Delta \Nx = 1$ while holding
        $\Ntot$ fixed.
        For free-streaming radiation (left), the two effects each reduce the amplitude at
        recombination (indicated by the vertical black line), while with fluidlike radiation (right)
        their effects partially cancel.
    }
    \label{fig:pressure_supp_vs_f_fs_effects}
\end{figure}

The differing impacts on the first peak compared to larger $\ell$ derive from the transition to
matter domination shortly before recombination, as the modes that contribute to it entered the horizon after equality and do not complete a full oscillation before recombination.
As explained in Refs.~\cite{Hou:2011ec,Ge:2022qws} and reviewed in \cref{sec:vary_omega_r}, decreasing the fraction of pressure-supported matter suppresses the anisotropies on large scales but boosts them at $\ell \gtrsim 400$.
\Cref{fig:pressure_supp_vs_f_fs_effects} illustrates the interplay of this effect and the increase
(decrease) in the free-streaming fraction $\ffs$ under additional free-streaming (fluidlike)
radiation.
Specifically, \cref{fig:pressure_supp_vs_f_fs_effects} depicts the differences in the dynamics of
the Sachs-Wolfe term $\Theta_0 + \psi$ for various cosmologies (where $\Theta_0$ is the monopole of
the photon distribution and $\psi$ the gravitational potential~\cite{Ma:1995ey}), for wave number
$k = 0.016~\mathrm{Mpc}^{-1}$, which contributes significantly to first acoustic peak.
While the change to the pressure-supported matter fraction reduces the Sachs-Wolfe term at
recombination in both cases, the differences incurred by changing $\ffs$ have opposite sign.
Both effects are similar in magnitude.
When $\Delta \Nfs > 0$, these two effects conspire together to reduce the first peak, while in the
cosmology with additional $\Nfld$ the effects largely cancel.

In summary, the combined effects of changing the radiation density and free-streaming fraction yield
degeneracies in CMB anisotropies that are qualitatively distinct when considering additional
free-streaming or fluidlike radiation.
Free-streaming radiation results in a reduction in amplitude that is more coherent across scales but
with substantial oscillations about the overall shift.
Fluidlike radiation instead incurs a boost in amplitude at $\ell \gtrsim 400$ that is greater in
magnitude but has nearly negligible oscillations.
In this case, however, the first peak is largely unaffected.
The relative amplitude of the first and higher acoustic peaks thus plays a crucial role in breaking
the degeneracy between $\Nfld$, $A_s$, and $\YHe$.
In \cref{sec:diff_multipole_ranges}, we explore the manner in which low-$\ell$ data break this
degeneracy by comparing constraints from mock, \Planck{}-like data over all scales and restricted to
small scales.

\subsection{Breaking degeneracies with large scale structure}
\label{sec:degen_breaking}

\Cref{sec:vary_omega_r,sec:vary-ffs,sec:fs_vs_fld} focus on searches for signatures of light relics in primary CMB anisotropies because of their unique sensitivity to the dynamics of spatial perturbations in the radiation epoch.
But both the total radiation density (\cref{sec:vary_omega_r}) and the free-streaming fraction
(\cref{sec:vary-ffs}) remain degenerate to some degree with other cosmological parameters. These
degeneracies can be broken by other observations, like the light element abundances and large-scale
structure.
The former---namely, astrophysical measurements of the primordial helium fraction, $\YHe$---break degeneracies in the damping tail with the total density $\Ntot$.
In the remainder of this section, we discuss the role large-scale structure observables, such as CMB lensing and the matter power spectrum, can play in breaking these degeneracies.

\begin{figure}[t]
    \centering
    \includegraphics[width=\textwidth]{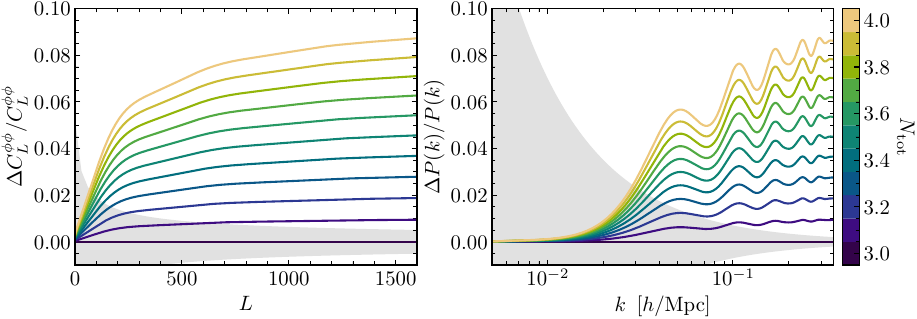}
    \caption{
        Impact of varying radiation density $\omega_r$ on the CMB lensing spectrum $C_\ell^{\phi\phi}$ (left) and the linear matter power spectrum $P_{mm}(k)$ (right) at fixed $\ffs$ and
        $\theta_D$ (by varying $\YHe$).
        Other $\LambdaCDM$ parameters are adjusted to fix $\theta_{s}$, $\aeq$, and the initial power in the mode that crosses the horizon at recombination (with wave number
        $k_{s, \star}$), as in \cref{fig:vary_omega_r_fix_theta_s,fig:vary_omega_r_fix_theta_s_theta_d}.
        Residuals are computed relative to a cosmology where $\Ntot = 3$.
        The grey bands depict cosmic variance for CMB lensing with $\fsky = 0.8$ binned with $\Delta \ell = 30$ (left) and for a survey volume of $20~\mathrm{Gpc}^3$ (right).
    }
    \label{fig:vary_omega_r_fix_theta_s_theta_d_Pk}
\end{figure}

We first consider variations in $\Ntot$ at fixed $\ffs$, as studied in \cref{sec:tilt_degen}, again
adjusting cosmological parameters to fix $\theta_s$, $\aeq$, and $\theta_D$ (via $\YHe$ as in
\cref{sec:helium}).
The remaining physical effect, the variation in the fraction of matter with pressure support
($\omega_b / \omega_{cb}$), increases power at high $\ell$
(\cref{fig:vary_omega_r_fix_theta_s_theta_d}), where the plasma oscillates in a deeper potential
well (due to the increased relative importance of CDM).
The same dynamics amplify large-scale structure, i.e., perturbations of the metric potentials and
the matter density, as evident in \cref{fig:vary_omega_r_fix_theta_s_theta_d_Pk}.
The relative increase in the CMB lensing and matter power spectra is around four times greater than
that in the temperature and polarization spectra at large $\ell$, which was at most $2\%$ over the
same range of $\Ntot$ (see \cref{fig:vary_omega_r_fix_theta_s_theta_d}).
These observables can therefore independently break the amplitude degeneracy discussed in
\cref{sec:f_fs_vs_pressure_supp}.

\begin{figure}
    \includegraphics[width=\textwidth]{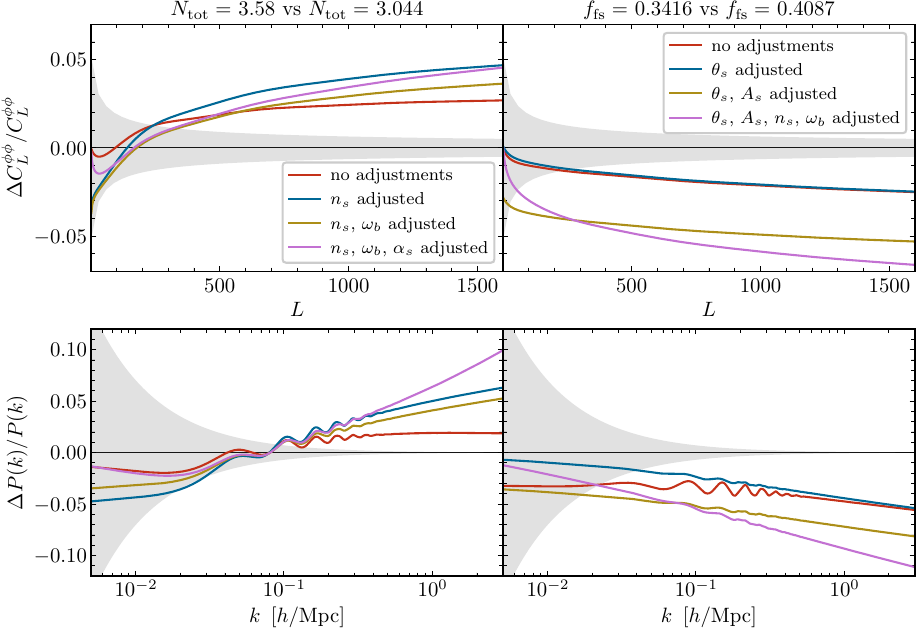}
    \caption{
        Relative changes to the CMB lensing potential power spectrum (top) and matter power spectrum (bottom) in cosmologies varying parameters along the
        tilt (left) and shift (right) degeneracies, comparing models with $\Ntot = 3.58$ and $3.044$
        at fixed $\ffs$ (left) and models with $\ffs = 0.34$ and $0.41$ at fixed $\Ntot$ (right).
        Parameters are varied according to \cref{eqn:tilt-degeneracy} (left) and \cref{eqn:shift-degeneracy} (right) as indicated on the legend; the CMB spectra for the corresponding cosmologies are displayed in \cref{fig:tilt_degen} and \cref{fig:shift_degen}, respectively.
        The grey bands depict cosmic variance as in \cref{fig:vary_omega_r_fix_theta_s_theta_d_Pk}.
    }
    \label{fig:shift_tilt_Pk_pp}
\end{figure}

We next explore to what extent large-scale structure observations can independently break the tilt (\cref{sec:tilt_degen}) and shift (\cref{sec:shift_degen}) degeneracies.
\Cref{fig:shift_tilt_Pk_pp} presents the relative changes to the CMB lensing power spectrum and matter power spectrum in the same
cosmologies considered in \cref{fig:tilt_degen,fig:shift_degen}, which display the changes to the
CMB temperature and polarization spectra.
We compare cosmologies with $\Ntot = 3.58$ and $\ffs = 0.3416$ to $\LambdaCDM$ for the tilt and shift
degeneracies, respectively.
None of the adjustments prescribed by the tilt degeneracy [\cref{eqn:tilt-degeneracy}] abate the
$\sim 3-4\%$ increase in power at $L \gtrsim 200$ in the lensing power spectrum incurred by the
increase in $\Ntot$ at fixed $\aeq$.
In fact, the parameter adjustments of \cref{eqn:tilt-degeneracy} increase the discrepancies at
smaller scales in both CMB lensing and the matter power spectrum, albeit only moderately on linear scales.
In contrast, both observables are less sensitive to changes in the free-streaming fraction,
all else equal.
However, adjusting $A_s$ and $n_s$ per the shift degeneracy [\cref{eqn:shift-degeneracy}] yield a pronounced impact on residuals, with differences again reaching the $5\%$ level in the linear regime when all shift degeneracy parameters are adjusted.
\Cref{fig:shift_tilt_Pk_pp} shows that, in addition to CMB lensing data, the full shape of the matter power spectrum could independently break the tilt and shift degeneracies deriving from current CMB data, along the lines of past work that considered self-interacting neutrinos~\cite{He:2023oke,Camarena:2023cku}.

\section{Constraints on dark radiation}
\label{sec:constraints}

We now apply the theoretical background of \cref{sec:background_cosmo} to interpret the impact of
the latest CMB and BAO data on parameter inference in dark-radiation scenarios.
We begin in \cref{sec:results-Ntot-ffs} by taking both $\Ntot$ and $\ffs$ as free parameters,
motivated by the discussion in \cref{sec:background_cosmo} which phrases the physical origin of
degeneracies with standard $\LambdaCDM$ parameters in terms of the total radiation density
$\omega_r$ and the fraction $\ffs$ thereof that freely streams.
In practice, for these models we vary the amount of both fluidlike and free-streaming radiation,
ignoring the SM prediction for the contribution from neutrinos.
To isolate the role of physical effects other than small-scale damping, we additionally study cases
that allow $\YHe$ to vary independently of its standard BBN prediction, as discussed in
\cref{sec:helium}.
We also consider varying the running of the scalar spectral index $\alpha_s$ (\cref{sec:alpha_s})
as an alternative means to compensate for radiation's impact on damping without altering dynamics
(as impacted by $\YHe$).
After studying general scenarios varying the total radiation density and its composition, in
\cref{sec:results-dNfs-dNfld} we consider models featuring additional light relics (on top of SM
neutrinos) that are either self-interacting ($\Delta \Nfld$) or free-streaming ($\Delta \Nfs$).
In particular, we apply the results of \cref{sec:results-Ntot-ffs} to understand any differential
impact of particular datasets on parameter inference in each case (as derives from their opposite
effects on the free-streaming fraction).
Finally, in \cref{sec:CMB-S4} we forecast parameter constraints from CMB-S4~\cite{CMB-S4:2016ple}
for all the aforementioned scenarios.
\Cref{tab:TTTEEE_constraints,tab:+lensing_constraints,tab:+act_constraints,tab:+desi_constraints} in
\cref{app:supplementary-results} tabulate constraints on other $\LambdaCDM$ parameters for a number
of the models and dataset combinations we consider.

To carry out the analyses in \cref{sec:results-Ntot-ffs,sec:results-dNfs-dNfld} (and those presented in \cref{fig:tilt_degen_mcmc,fig:shift_degen_mcmc}), we use the Boltzmann
code \class{} 3.2.3 interfaced with \cobaya{} 3.5.4~\cite{Torrado:2020dgo,2019ascl.soft10019T}.
We employ CMB temperature, polarization, and lensing data from \Planck{}, considering both 2018 data
via the nuisance-marginalized likelihoods~\cite{Planck:2018vyg, Planck:2019nip} (referred to as
``PR3'') and subsequent reanalyses (``PR4'').
Specifically, the latter uses the \texttt{CamSpec} likelihood~\cite{Rosenberg:2022sdy} for data at
$\ell \geq 30$, \texttt{LoLLiPoP} for $E$-mode data~\cite{Tristram:2023haj} and \texttt{Commander}
for temperature data~\cite{Planck:2019nip} at $\ell < 30$ (the latter of which employs 2018 data),
and the PR4 lensing likelihood from Ref.~\cite{Carron:2022eyg}.
The combinations PR3 and PR4 each include their respective lensing datasets unless otherwise
specified, and in some cases we use the combined ACT DR6/\Planck{} PR4 lensing
dataset~\cite{ACT:2023dou, ACT:2023kun}.
Finally, we also study the impact of BAO observations either from DESI's first data
release~\cite{DESI:2024lzq,DESI:2024mwx,DESI:2024uvr} or from the Sloan Digital Sky Survey (SDSS),
the latter including the Main Galaxy Sample from SDSS DR7~\cite{Ross:2014qpa}, the Baryon
Oscillation Spectroscopic Survey (BOSS) DR12 galaxies~\cite{BOSS:2016wmc}, and the Extended BOSS
(eBOSS) DR16 luminous red galaxies~\cite{eBOSS:2020lta, eBOSS:2020hur, eBOSS:2020yzd}.\footnote{
    In principle, the phase shift due to free-streaming radiation also shifts the BAO
    peak~\cite{Baumann:2017lmt,Baumann:2017gkg,Green:2020fjb}, an effect measured in SDSS~\cite{Baumann:2019keh} and DESI~\cite{Whitford:2024ecj} data.
    However, the template fitting methods used to measure the drag scale are relatively robust to changes in physics before recombination, even in extensions to $\LambdaCDM$ that include additional light relics~\cite{Bernal:2020vbb}.
    We therefore use BAO likelihoods without modification, as is standard~\cite{DESI:2024mwx, DESI:2024ude}.
}

We sample posteriors using the \cobaya{} implementation of the Metropolis-Hastings method for Markov
chain Monte Carlo~\cite{Lewis:2013hha}.
Denoting a uniform distribution between $a$ and $b$ as $\mathcal{U}(a, b)$, we take flat priors for $\LambdaCDM$ parameters: $100\theta_s\sim\mathcal{U}(0.5, 10)$, $\omega_b\sim\mathcal{U}(0.005, 0.1)$, $\omega_c\sim\mathcal{U}(0.001, 0.99)$, $\ln \left(10^{10}A_s \right)\sim\mathcal{U}(1.61, 3.91)$, $n_s\sim\mathcal{U}(0.8, 1.2)$, and $\taureio\sim\mathcal{U}(0.01, 0.8)$.
The total amount of radiation and the free-streaming fraction are sampled via
$\Ntot \sim \mathcal{U}(2.0, 4.5)$ and the ratio $\Nfld/\Ntot \sim \mathcal{U}(0, 1.0)$.
When not otherwise fixed, we take $\YHe \sim \mathcal{U}(0.01, 0.5)$ and
$\alpha_s \sim \mathcal{U}(-0.05, 0.05)$.
For the cases considered in \cref{sec:results-dNfs-dNfld}, we take $\Delta \Nfs$ and
$\Delta \Nfld \sim \mathcal{U}(0, 1)$.
We ensure that our presented results are robust to both thinning and the amount of samples dropped
as burn-in.
All posteriors presented contain between $4,000$ and $11,000$ independent samples, and we slightly smooth posterior contours to improve legibility without modifying their shape.

\subsection{Impact of recent data on radiation density and composition}\label{sec:results-Ntot-ffs}

\begin{figure}[t]
    \centering
    \includegraphics[width=\textwidth]{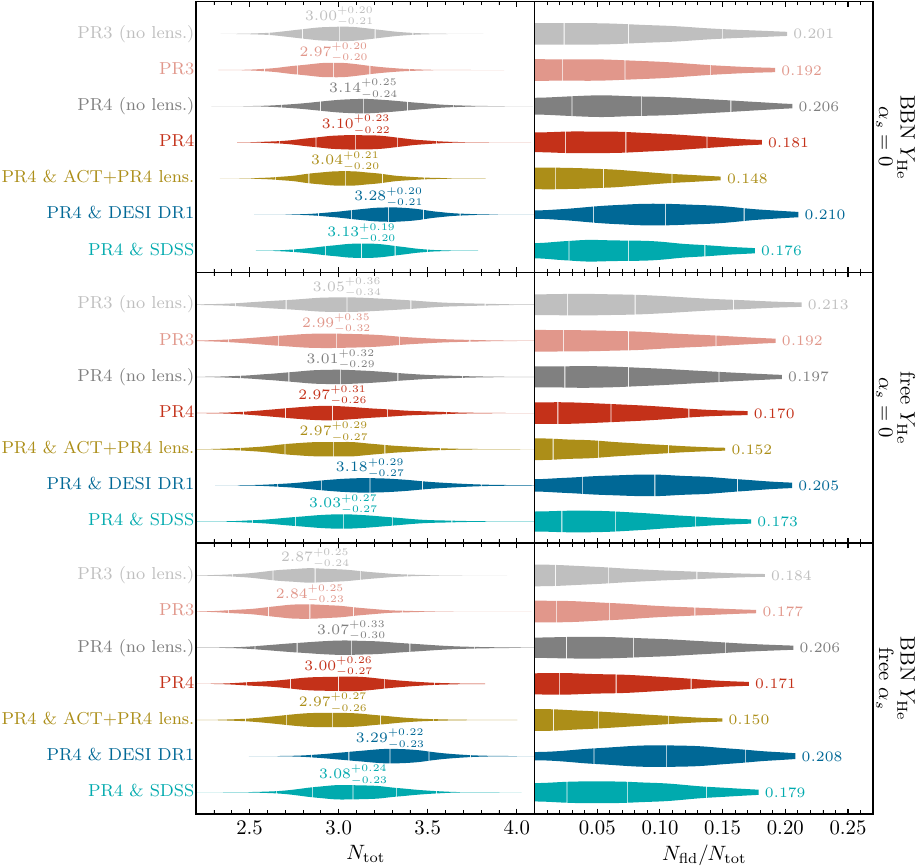}
    \caption{
        Posterior distributions for $\Ntot$ and $\Nfld/\Ntot$ for models with $\Ntot$ and $\ffs$ free.
        Results are derived using \Planck{} PR3 and PR4 CMB data, with or without lensing, and in
        combination with BAO data from DESI and SDSS or lensing data from ACT DR6 as labeled.
        The primordial helium yield $\YHe$ is fixed to the BBN prediction in the top panels and
        independently varied in the middle panels.
        The bottom panels marginalize over the running of the spectral tilt $\alpha_s$, again fixing $\YHe$ to the BBN prediction.
        Posteriors on the left are labeled by their median and $1\sigma$ quantiles, with corresponding white lines indicating the median, $1\sigma$, and $2\sigma$ contours.
        On the right, posteriors are truncated at the 95th percentile (whose value is also labeled); vertical white lines indicate the median and $\pm 1\sigma$ quantiles.
    }
    \label{fig:Ntot-chi-violin}
\end{figure}

We first analyze the impact of recent datasets on constraints for models where both the radiation
density and the free-streaming fraction vary, isolating what features of recent data are responsible
for changes in parameter inference.
Later (in \cref{sec:results-dNfs-dNfld}) we apply this analysis to explain differences in
constraints on additional free-streaming radiation or fluidlike radiation beyond the SM prediction.
\Cref{fig:Ntot-chi-violin} summarizes measurements on $\Ntot$ and $\ffs$ for various dataset
combinations in scenarios that fix $\YHe$ to its BBN prediction, vary $\YHe$ independently, and
vary the running of the spectral index, $\alpha_s$.

\subsubsection{\Planck{} PR3 versus PR4}\label{sec:pr3-vs-pr4}

\begin{figure}[t]
    \centering
    \includegraphics[width=3in]{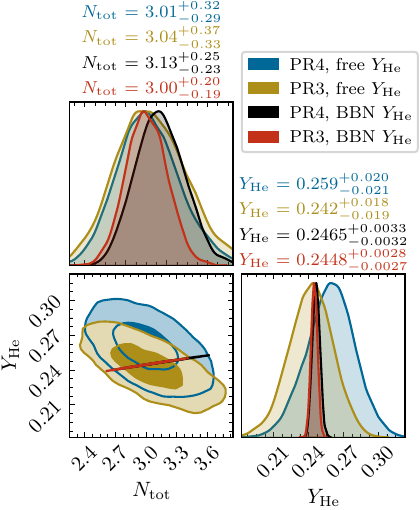}
    \includegraphics[width=3in]{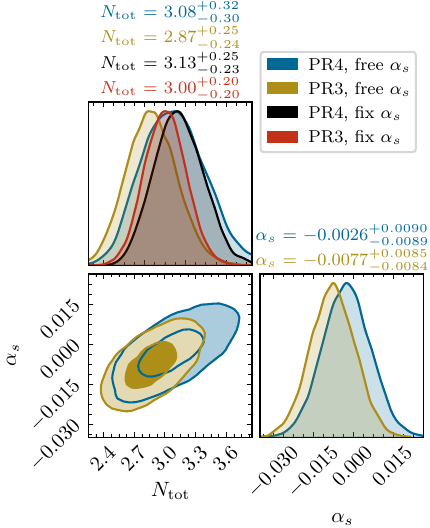}
    \caption{
        Posterior distributions for models where $\Ntot$ and $\ffs$ both vary, comparing results
        using the \Planck{} PR3 and PR4 temperature and polarization data (excluding lensing data).
        In the left figure, $\YHe$ is either independently varied (blue and gold for PR4 and PR3,
        respectively) or fixed to the BBN prediction (black and red).
        The right figure compares results that also vary $\alpha_s$ (blue and gold) to those
        that fix it to zero (black and red), both with $\YHe$ set to its BBN prediction.
        Diagonal panels depict kernel density estimates of one-dimensional posterior distributions
        normalized relative to their peak value to facilitate comparison.
        The median and corresponding $\pm 1 \sigma$ uncertainties for each parameter are reported
        above the diagonal panels.
        The off-diagonal panels display the 1 and $2 \sigma$ contours of the two-dimensional joint
        posterior density for pairs of parameters (i.e., the $39.3\%$ and $86.5\%$ mass levels).
    }
    \label{fig:pr3_vs_pr4_YHe_alpha}
\end{figure}

The PR4 and PR3 data releases largely yield consistent inferences of $\LambdaCDM$ parameters; in
general, the PR4 release reduces parameter uncertainties by $10\%$ to
$20\%$~\cite{Tristram:2023haj,Rosenberg:2022sdy}.
However, PR4 prefers slightly higher $\Nfs = 3.08 \pm 0.17$~\cite{Rosenberg:2022sdy} compared to PR3
($2.92 \pm 0.19$~\cite{Planck:2018vyg}), a result marginally closer to the SM prediction for $\Nfs =
3.044$ for neutrinos.
This preference for additional radiation persists in models where $\Ntot$ and $\ffs$ both vary, as
shown in \cref{fig:pr3_vs_pr4_YHe_alpha}.
When fixing $\YHe$ to its BBN prediction (as a function of $\Ntot$ and the baryon-to-photon ratio),
the posteriors over $\Ntot$ from PR4 shift around half a standard deviation higher and also broaden
marginally.
When $\YHe$ is instead taken as a free parameter, the posteriors are less offset in $\Ntot$ and the
PR4 measurement is slightly more precise than PR3's.
Notably, however, PR4 in general prefers higher values of $\YHe$ regardless of the value of $\Ntot$.
Comparing the two-dimensional posteriors that do and do not assume the prediction from BBN, for
which $\YHe$ increases with $\Ntot$, suggests that PR4's preference for larger $\Ntot$ when assuming consistency with BBN is driven by its preference for larger values of $\YHe$.
This comparison also suggests that PR3 data is modestly more consistent with SM predictions for $\YHe$.
Moreover, the $\Ntot$ posteriors for PR4 and PR3 both center at $3$ when marginalizing over $\YHe$.
This suggests that the effects of changing $\YHe$ better explain the differences between the two
datasets than any effect deriving specifically from the radiation content.

\begin{figure}[t]
    \centering
    \includegraphics[width=\textwidth]{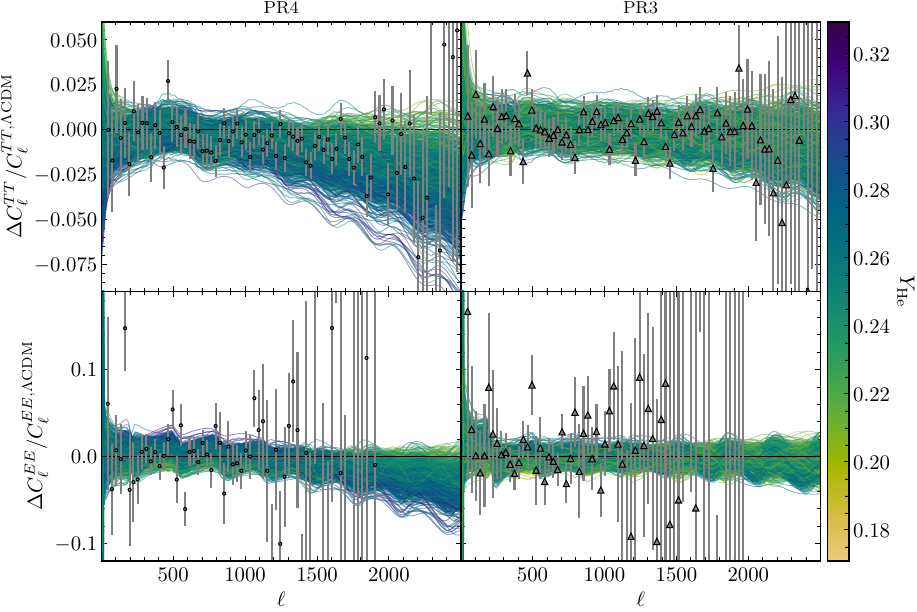}
    \caption{
        Residuals of the CMB temperature and polarization spectra relative to the $\LambdaCDM$ best fit (derived from PR3 data) for samples from posteriors with $\Ntot$, $\ffs$, and $\YHe$ as free parameters.
        The left and right panels use PR4 and PR3 temperature and polarization data, respectively,
        both excluding lensing data.
        Curves are colored by each sample's value of $\YHe$; PR4 and PR3 data (binned in interval
        $\Delta \ell = 30$) and standard deviations appear as grey circles and triangles,
        respectively.
    }
    \label{fig:free_f_fs_free_YHe_spag}
\end{figure}

To investigate the origin of PR4's preference for larger values of $\YHe$,
\cref{fig:free_f_fs_free_YHe_spag} displays the residuals of CMB spectra for 1000 posterior samples
for both PR4 and PR3 (i.e., those presented in \cref{fig:pr3_vs_pr4_YHe_alpha} with $\Ntot$, $\ffs$
and $\YHe$ free), each colored by its value of $\YHe$.
These results indicate that PR4 prefers both excess damping in temperature at high $\ell$ and a
higher polarization amplitude at $100 \leq \ell \leq 1000$ (driven by several data points in this
interval skewing several standard deviations above the best-fit prediction and their PR3
counterparts).
Both features can be better fit by increasing $\YHe$, deriving from the incurred slowing of
recombination.
Namely, with slower recombination the quadrupole has more time to radiate before last scattering,
generating a larger amplitude of polarization~\cite{Zaldarriaga:1995gi, Hu:1997hv,
Baryakhtar:2024rky}.
A broader visibility function also increases the degree of so-called Landau damping on small
scales~\cite{Zaldarriaga:1995gi, Weinberg:2008zzc}.

\Cref{fig:pr3_vs_pr4_YHe_alpha} also displays results that vary all of $\Ntot$, $\ffs$, and
$\alpha_s$.
As noted in \cref{sec:alpha_s}, the running of the spectral tilt can mimic the effect of small-scale
damping by changing initial conditions rather than dynamics.
Marginalization over the running slightly weakens constraints on $\Ntot$ with either PR3 or PR4 data.
However, $\alpha_s$ cannot mimic the enhancement of the polarization amplitude from increasing $\YHe$
as preferred by PR4 data.
PR3 favors negative running more so than PR4, as attributed to mild tensions between low and high
multipoles in $\LambdaCDM$~\cite{Planck:2018vyg}, correlating to a slightly lower radiation density.
In contrast to results for $\LambdaCDM$ parameters, PR4 yields slightly weaker constraints than PR3
on $\alpha_s$ (as well as on $\Ntot$ and $\YHe$), in curious contradiction to its greater data
volume and multipole range.
This finding may derive from the aforementioned internal inconsistencies in PR3
data~\cite{Planck:2018vyg} that may be mitigated in PR4, a possibility that warrants further
investigation.

\subsubsection{Role of CMB lensing data}

As was true with the \Planck{} PR3 data release~\cite{Planck:2018vyg, Blinov:2020hmc}, incorporating
PR4 lensing data does not qualitatively impact parameter inference for models that vary both $\Ntot$
and $\ffs$, as shown in \cref{fig:Ntot-chi-violin}.
\Planck{}'s CMB lensing data lack the precision to meaningfully break parameter degeneracies in the manner discussed in \cref{sec:degen_breaking}.
ACT's DR6 lensing data also have little impact on the inference of $\Ntot$, regardless of what additional parameters are varied ($\YHe$ or $\alpha_s$).
However, ACT lensing data (in combination with PR4) do have a noticeable impact on the fraction of
radiation that is fluidlike.

\begin{figure}[t]
    \includegraphics[width=\textwidth]{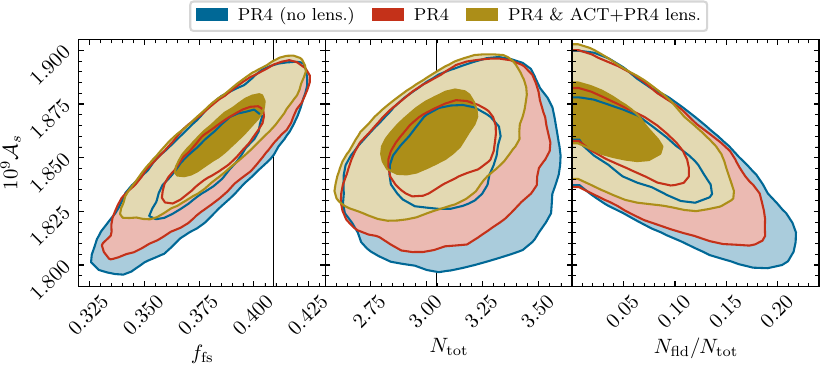}
    \caption{
        Joint posteriors over $\Ntot$, $\Nfld/\Ntot$, and $\mathcal{A}_s = \Astau$ for
        models varying both $\Ntot$ and $\ffs$, fixing $\YHe$ to BBN predictions and using PR4 CMB data without lensing (blue), with PR4 lensing (red), and with PR4+ACT DR6 lensing (gold).
        The measurement of the amplitude of structure from CMB lensing data partly break the
        degeneracy between $\mathcal{A}_s$ and $\ffs$, tightening upper limits on
        $\Nfld / \Ntot$; the marginal posteriors over $\Ntot$ are less affected.
        These findings also hold when $\alpha_s$ or $\YHe$ freely vary.
    }
    \label{fig:lensing_effects}
\end{figure}

The increase in constraining power with ACT lensing data derives from effects of varying the
free-streaming fraction---in particular, by better measuring $A_s$ and breaking its degeneracy with
$\ffs$ (see \cref{sec:degen_breaking}).\footnote{
    Note that the impact of CMB lensing data depends sensitively on the inference of the optical
    depth from low-$\ell$ polarization data, which is required to determine $A_s$ from
    $\mathcal{A}_s = \Astau$.
}
\Cref{fig:lensing_effects} compares posteriors from PR4 that exclude and include lensing
data from PR4 and from the combination of ACT DR6 and PR4.
Decreasing the free-streaming fraction increases the amplitude of the CMB temperature and
polarization spectra, requiring a reduced $\mathcal{A}_s$ to fit the data
(\cref{sec:fs_vs_fld}).
However, moving along the shift degeneracy exacerbates differences in the lensing potential (see \cref{fig:shift_tilt_Pk_pp}).
Therefore, the additional lensing information disfavors only the lower values of $\mathcal{A}_s$ allowed by the primary CMB.
Since models with additional $\Nfld$ radiation prefer lower values of $A_s$, the addition of the PR4 + ACT DR6 lensing likelihood reduces the fraction of
radiation that can be fluidlike, as is seen in the posteriors over $\ffs$ and $\Nfld/\Ntot$ in
\cref{fig:lensing_effects}.
Upper limits on $\Ntot$ also decrease slightly, since the possibility of adding radiation in
fluidlike form is more limited.

\subsubsection{Preference for additional radiation with BAO data}
\label{sec:Nfld-BAO}

We now study the impact of low-redshift distance measurements from baryon acoustic oscillations.
In models (like those we consider here) whose late-time expansion histories are well described by
the flat $\LambdaCDM$ model, low-redshift distance measurements only contribute meaningfully in
models with some degree of geometric degeneracy in the distance to last scattering: within the
standard six-parameter $\LambdaCDM$ model, \Planck{}'s measurements of $\theta_s$ and $\omega_{cb}$
indirectly constrain the late-time expansion history better than do direct measurements from BAO or
supernovae distances.
These data, however, are crucial in scenarios with geometric freedom in the CMB---for example, when
taking the neutrino mass sum as a free parameter~\cite{Loverde:2024nfi}, in early recombination
scenarios~\cite{Baryakhtar:2024rky}, or in models departing from flat $\LambdaCDM$ at late times.

BAO distances are effectively parameterized by the matter fraction
$\Omega_m = \omega_m / (\omega_m + \omega_\Lambda)$ (which encodes the redshift dependence, i.e.,
when the dark-energy era began) and $\omega_m \rdrag^2$~\cite{Loverde:2024nfi}, where
$\omega_m = \omega_b + \omega_c + \omega_\nu$ is
the density in matter at late times\footnote{
    The distinction between $\omega_{cb}$ and $\omega_m$ is irrelevant because our analyses fix the
    neutrino mass sum and therefore fix $\omega_m$ as a function of $\omega_{cb}$.
} and $\rdrag$ the comoving sound horizon of the plasma at the time baryons
decouple~\cite{Eisenstein:1997ik}.
In dark radiation models, one would naively expect no such additional freedom when $\theta_s$ and
$\aeq$ are fixed, because then
$\omega_\Lambda \propto \omega_{cb} \propto \omega_r$~\cite{Baryakhtar:2024rky, Loverde:2024nfi}.
In this case, $\Omega_m$ is unchanged, and, because the sound horizon is inversely proportional to
$\sqrt{\omega_r}$ at fixed $\aeq$, $\omega_m \rdrag^2 \approx \omega_{cb} \rdrag^2$ is unchanged as well.
However, \cref{fig:tilt_degen_mcmc} shows that, due to the competing effect of changes to the
pressure-supported matter fraction, posteriors from \Planck{} do not precisely follow
$\omega_{cb} \propto \omega_r$---that is, the posteriors are driven by a compromise between fixing
$\omega_{cb} / \omega_r$ (which fixes the radiation-driving and early integrated Sachs-Wolfe
effects) and $\omega_b / \omega_{cb}$.
The competition between these two effects is further illustrated by \cref{fig:pressure_supp_mcmc} in \cref{app:supplementary-results}.
We now test the impact of BAO data in breaking the resulting additional geometric freedom.

\begin{figure}[t]
    \includegraphics[width=\textwidth]{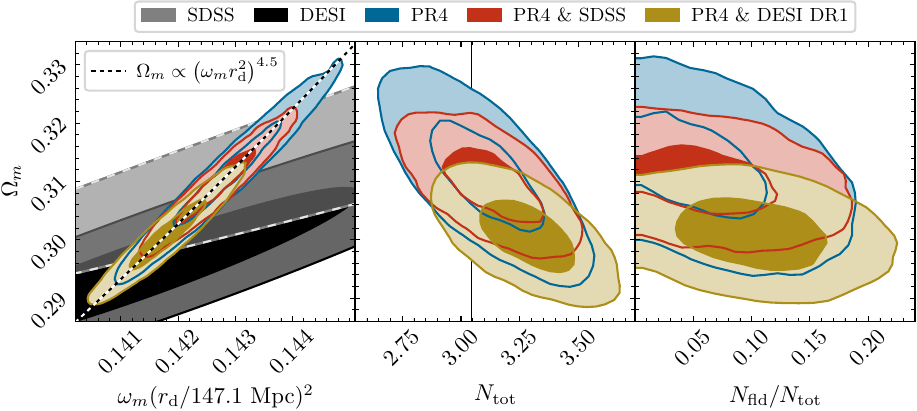}
    \caption{
        Joint posteriors over $\omega_m \rdrag^2$, $\Ntot$, $\Nfld/\Ntot$, and $\Omega_m$ for models
        varying $\Ntot$ and $\ffs$, using dataset combinations labeled in the legend.
        The posteriors deriving from SDSS (grey, dashed) or DESI (black) alone display constraints
        for the flat $\LambdaCDM$ model (following Ref.~\cite{Loverde:2024nfi}), and their overlap
        with the PR4 CMB-only posterior (blue) determines the extent of the joint posteriors (red
        and gold).
        Note that only one end of SDSS's $2 \sigma$ region is visible within the axes limits;
        since we only use a subset of eBOSS tracers, the joint posterior is centered at larger
        $\Omega_m$ and $\omega_m \rdrag^2$ than depicted.
    }
    \label{fig:BAO_effects}
\end{figure}

Interestingly, \cref{fig:Ntot-chi-violin} shows that DESI data push posteriors toward greater values
of $\Ntot$ across all model extensions.
DESI BAO data prefer a lower matter fraction (at any given $\omega_m \rdrag^2$) compared to
\Planck{} data alone, as evident in the posteriors in \cref{fig:BAO_effects}.
The CMB's geometric degeneracy is clear in the joint posterior over $\Omega_m$ and
$\omega_m \rdrag^2$ in \cref{fig:BAO_effects}; analytically, fixing $\theta_s$ in flat $\LambdaCDM$
cosmologies approximately requires~\cite{Loverde:2024nfi}\footnote{
    Note that a numerical fit derived from the PR4-only posteriors yields
    $\Omega_m \propto \left( \omega_m \rdrag^2 \right)^{4.5}$, as displayed in dashed black in \cref{fig:BAO_effects}.
    The deviation in exponent from that in \cref{eqn:theta_s_Omega_m_degen} is due to correlations of $\theta_s$ with $\omega_b$ and $\omega_c$ as well as the free-streaming fraction $\ffs$, none of which are accounted for in the approximations that yield \cref{eqn:theta_s_Omega_m_degen}.
}
\begin{align}
    \label{eqn:theta_s_Omega_m_degen}
    \left. \Omega_m \right\vert_{\theta_s}
    \propto \left( \omega_m \rdrag^2 \right)^{5}.
\end{align}
Because $\omega_m$ increases sublinearly with $\omega_r$ (see \cref{fig:tilt_degen_mcmc,fig:pressure_supp_mcmc}), the matter
fraction can be reduced (as preferred by DESI) when increasing $\omega_r$ while still satisfying the
degeneracy direction in \cref{eqn:theta_s_Omega_m_degen}.
BAO data from SDSS do not drive the same preference for larger $\omega_r$, as also evident in
\cref{fig:Ntot-chi-violin}, because the SDSS data prefer a larger matter fraction at any
$\omega_m \rdrag^2$ than DESI, closer to those PR4 infers in $\LambdaCDM$.
Note that Ref.~\cite{Loverde:2024nfi} showed that DESI's preference for lower $\Omega_m$ than SDSS
derives entirely from the two measurements that are most in tension with those from SDSS.

To maintain concordance with CMB data, the preference for additional radiation in constraints that include the DESI DR1 BAO data is mostly accomplished through the addition of fluidlike radiation.
Since the CMB prefers $\ffs$ close to the $\LambdaCDM$ value and $\Nfld$ affects $\ffs$ less dramatically than additional free-streaming radiation [per \cref{eqn:perturb_ffs}] the posteriors for $\Nfld/\Ntot$ shift to higher values (\cref{fig:Ntot-chi-violin}) across all model extensions when including DESI data.

\subsection{Constraints on interacting and noninteracting dark radiation}\label{sec:results-dNfs-dNfld}

\begin{figure}[t]
    \centering
    \includegraphics[width=\textwidth]{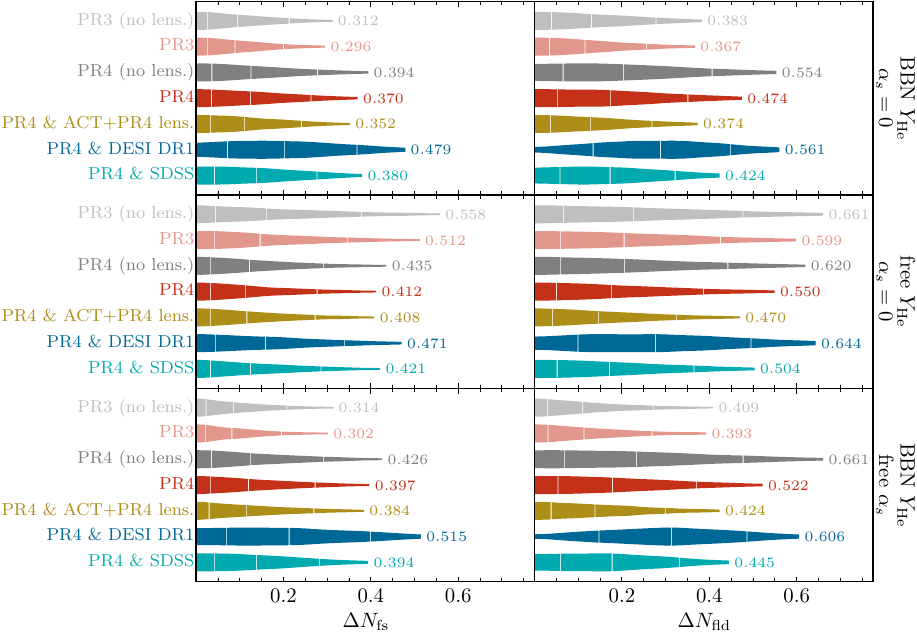}
    \caption{
        Marginal posterior distributions over $\Delta \Nfs$ and $\Delta \Nfld$ for models adding either free-streaming or fluidlike dark radiation.
        Results are presented as in \cref{fig:Ntot-chi-violin}.
    }
    \label{fig:dNfs-dNfld-violin}
\end{figure}

We now turn to models that strictly allow radiation in addition to the contributions from SM neutrinos of $\Ntot = \Nfs = 3.044$, parametrized by $\Delta \Nfs$ and $\Delta \Nfld$.
We interpret these results as special cases of \cref{sec:results-Ntot-ffs}, which allowed for arbitrary amounts of free-streaming and fluidlike radiation, as varying $\Delta \Nfs$ or $\Delta \Nfld$ prescribes a specific relationship between the total density $\omega_r$ and the free-streaming fraction $\ffs$.
We continue to examine the impact of all three model extensions considered above: $\YHe$ fixed to the BBN prediction, $\YHe$ allowed to freely vary, and $\alpha_s$ allowed to vary.
The marginal posterior distributions for these models are reported in \cref{fig:dNfs-dNfld-violin}.
Consistent with our findings for models with $\Ntot$ and $\ffs$ both free, PR4 allows for slightly
more radiation than PR3 across all model extensions, deriving from differences in the polarization
amplitude that drive a preference for higher helium yield (see \cref{fig:free_f_fs_free_YHe_spag}).

Like the case with $\Ntot$ and $\ffs$ both free, the combination of PR4 and ACT DR6 lensing data has
a substantially greater impact on upper limits than the PR4 lensing data alone.
In particular, lensing has the greatest impact on $\Delta \Nfld$.
As mentioned in \cref{sec:fs_vs_fld}, additional fluidlike radiation decreases the free-streaming fraction [per \cref{eqn:perturb_ffs}], which boosts the amplitude of both the temperature and polarization power spectra.
In turn, lower values of $\mathcal{A}_s$ are required to fit the data.
However, lensing data drive a preference for higher values of $\mathcal{A}_s$ (\cref{fig:lensing_effects}), further restricting $\Delta \Nfld$.
Even when marginalizing over $\YHe$ and $\alpha_s$, the addition of ACT lensing data significantly improves upon PR4 in limits on $\Delta \Nfld$.
In contrast, constraints from PR3 alone are stronger than PR4 and ACT DR6 constraints when $\YHe$ is consistently set to the BBN predictions, driven by PR3 data's independent preference for a lower radiation density (see \cref{sec:pr3-vs-pr4}).

The preference for additional radiation when including DESI BAO data persists in models with $\Delta \Nx$, regardless of variations in $\YHe$ and $\alpha_s$.
BAO data from SDSS, on the other hand, have little effect because its geometric preferences are more consistent with \Planck{}'s.
DESI data drive a more substantial preference for $\Delta \Nfld$ than $\Delta \Nfs$ since the former's effects in the CMB are less distinctive (see \cref{sec:fs_vs_fld}).
This preference for a greater radiation density (and therefore a smaller sound horizon) in turn
pushes $H_0$ to larger values
$69.8^{+1.2}_{-1.1}~\mathrm{km}/\mathrm{s}/\mathrm{Mpc}$ (see \cref{app:supplementary-results}) as
reported in Ref.~\cite{Allali:2024cji}.
Models with fluidlike dark radiation that take the BBN predictions for $\YHe$ have a median value of $H_0$ value $2.1 \sigma$ away from that from SH0ES~\cite{Riess:2021jrx}, while in the free-streaming case the tension is $2.7\sigma$.
Allowing $\YHe$ and $\alpha_s$ to vary does not significantly improve the agreement of $H_0$ to the local measurement.

\subsection{Forecasts for CMB-S4}\label{sec:CMB-S4}
The planned CMB-S4 experiment is expected to improve constraints on $\Nfs$ by almost an order of magnitude.
In this section, we forecast measurements for CMB-S4 in the models studied above for the latest experimental configuration from the preliminary baseline design report~\cite{CMB-S4:pbdr}.
In particular, we use noise curves for the latest specification of the S4-wide portion of the experiment which involves two large aperture telescopes located in Chile.
The noise curves are constructed using the dark radiation anisotropy
flowdown team (DRAFT) tool, which models both galactic and extragalactic foregrounds~\cite{DRAFTtool,Raghunathan:2023yfe,SPTpol:2025kpo,Panexp_2025,Zonca_2021,Thorne_2017}.\footnote{See \href{https://github.com/sriniraghunathan/DRAFT/blob/master/products/202310xx_PBDR_config/s4wide_202310xx_pbdr_config/}{here} for the noise curves used.}
We adopt a galactic mask that retains over 56\% of the sky and implement these forecasts with the \texttt{mock\_cmb\_likelihood} from \montepython{}~\cite{Audren:2012wb,Brinckmann:2018cvx}, taking multipoles with $ 30 \leq \ell \leq 5000$ into account.
We use the mean of the posteriors for $\LambdaCDM$ parameters from the \Planck{} PR4 dataset for the fiducial cosmology (see the first column of \cref{tab:+lensing_constraints}).
We also take a Gaussian prior over $\taureio$ with mean $0.0577$ and standard deviation $0.0062$ as provided by \Planck{} PR4 data, as CMB-S4 will not have access to the large-scale modes that directly constrain reionization.

We implement a joint likelihood including lensed temperature and polarization as well as reconstructed lensing spectra.
We find that, for the class of models considered here, unlensed spectra give nearly identical
results as lensed spectra, with the only difference being that unlensed spectra yield slightly more
precise measurements of $\theta_s$.\footnote{
    Delensing the primary anisotropies effectively removes the peak smearing caused by lensing, providing more precise peak location information and improving bounds on
    $\theta_s$~\cite{Green:2016cjr,Hotinli:2021umk}.
    Delensing and a more careful modeling of covariance are more important for physics related
    to late-time structure, such as massive neutrinos (which suppress structure growth).
}
These findings are consistent with prior Fisher forecasts for $\Nfs$~\cite{Green:2016cjr}, which
further only a small improvement from delensing the primary anisotropies.
Our use of lensed spectra is thus more conservative than using spectra that are delensed or fully
unlensed (i.e., the idealized limit of perfect delensing).
We also neglect the non-Gaussian contributions to the covariance induced by lensing, which
Ref.~\cite{Green:2016cjr} found had little effect on $\Nfs$ forecasts.
Although these non-Gaussian contributions have been shown to lead to overly optimistic constraints on $A_s$, we find that our forecasts are unaffected, despite the interesting interplay between $A_s$ and fluidlike radiation (\cref{sec:f_fs_vs_pressure_supp}).
We leave a proper treatment of delensing and covariance in CMB-S4 forecasts to future work.

CMB-S4's design sensitivity requires accurate modeling of nonlinear structure growth.
At $\ell = 1500$ in the lensing potential spectra, the linear and nonlinear power spectra differ by nearly $20\%$, which CMB-S4 will easily differentiate between.
To model these nonlinearities, we use \texttt{HMCode}-2016~\cite{Mead:2016zqy} with \texttt{nonlinear\_min\_k\_max} set to $5~\mathrm{Mpc}^{-1}$.\footnote{We find that increasing \texttt{nonlinear\_min\_k\_max} to larger values does not substantially change the lensing potential spectra relative to CMB-S4's sensitivity.}
Although \texttt{HMCode} is calibrated to simulations that do not include additional light relics, we find that it gives similar results to \texttt{CLASS\_PT}~\cite{Chudaykin:2020aoj}, which consistently treats differences to matter perturbations due to the additional dark radiation within standard perturbation theory at one loop.
Since the differences between the two nonlinear codes is only at the percent-level, we opt to use \texttt{HMCode} for simplicity.

\begin{figure}[t]
    \centering
    \includegraphics[width=\textwidth]{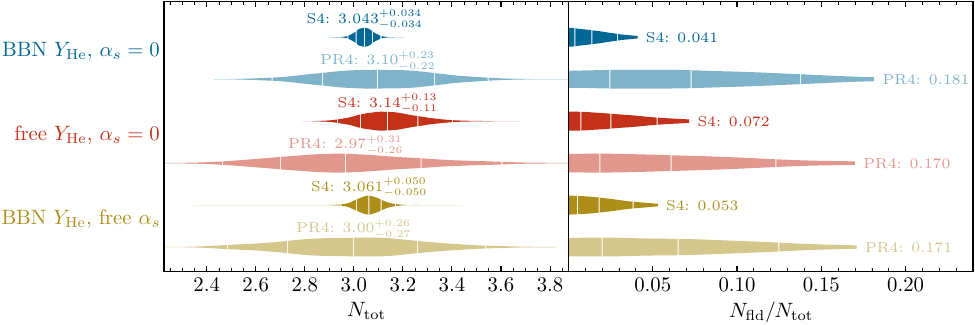}
    \includegraphics[width=\textwidth]{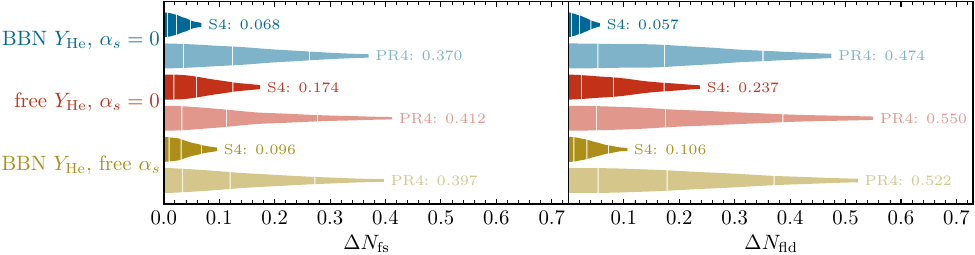}
    \caption{
        Forecasted constraints from CMB-S4 compared with current ones from the \Planck{} PR4 dataset for models with $\Ntot$ and $\ffs$ free (top) as well as models with additional free-streaming or fluidlike radiation (bottom).
        Both cases included lensing reconstruction data.
        Results are shown for models with $\YHe$ set to its BBN prediction, $\YHe$ freely varying, and $\alpha_s$ freely varying as labeled.
    }
    \label{fig:s4_vs_pr4_1d}
\end{figure}

We present forecasts for models with $\Ntot$ and $\ffs$ free, as well as models individually varying
$\Delta \Nfld$ and $\Delta \Nfs$, in \cref{fig:s4_vs_pr4_1d}.
When $\YHe$ is fixed to its BBN prediction, we observe a factor of seven improvement in sensitivity in $\Ntot$ between the \Planck{} PR4 dataset combination and CMB-S4, with the $1 \sigma$ intervals improving from $\sigma(\Ntot) = 0.23$ to $\sigma(\Ntot) = 0.034$.
The cases with $\YHe$ free show only a factor of three improvement in sensitivity. However, the $1 \sigma$ interval for $\YHe$ also improves by a factor of three; we forecast a measurement $\sigma(\YHe) \sim 0.008$ from CMB data alone with CMB-S4, independent of BBN (see \cref{fig:YHe-alpha_s-violin-all} in \cref{app:supplementary-results}.)
In models with $\alpha_s$ free, CMB-S4 yields a similar factor of six improvement for $\Ntot$ and a factor of $2.2$ for $\alpha_s$, improving from $\sigma(\alpha_s) = \num{8.4e-3}$ with PR4 data to $\sigma(\alpha_s) = \num{3.8e-3}$ with CMB-S4.

Interestingly, when $\YHe$ and $\ffs$ are free, the forecasted posteriors are not centered on the fiducial values of $\Ntot$ and $\YHe$.
Since we fix the photon density, the maximum value of $\ffs$ is determined by $\Ntot$---that is, the model can only increase $\ffs$ by also increasing $\Ntot$.
In the sense that $\ffs$ is an effective summary statistic of the physical features constrained by
the data, this constraint removes parameter space with lower $\Ntot$ and higher $\ffs$ that would
otherwise fit the data well.
Such a diagonal cut thus shifts the mode of the posterior.

For models with additional radiation on top of the minimum $\LambdaCDM$ prediction for $\Nfs$, CMB-S4 will improve sensitivity to both free-streaming and fluidlike radiation by nearly a factor of five when taking BBN predictions for $\YHe$.
The $95\%$ upper limit on additional free-streaming radiation improves from $\Delta \Nfs \leq 0.37$ to $\Delta \Nfs \leq 0.068$, while that for fluidlike radiation improves from $\Delta \Nfld \leq 0.47$ to $\Delta \Nfld \leq 0.057$.
\Cref{fig:s4_vs_pr4_1d} displays comparable improvements in sensitivity when $\YHe$ and $\alpha_s$ are free.

\section{Conclusions}
\label{sec:conclusion}

In this paper, we studied the cosmological signatures of the radiation content of the
early Universe---in particular, considering scenarios beyond the predictions for SM neutrinos that
feature light degrees of freedom that are strongly self-interacting, collisionless, or both in some
proportion.
In \cref{sec:background_cosmo}, we reviewed the physical origin of parameter degeneracies in the CMB
anisotropy spectra that arise in such scenarios, phrasing the relevant effects in terms of the total
radiation density $\omega_r$ and the fraction $\ffs$ thereof that freely streams (as SM neutrino do
after weak decoupling) in order to disentangle which parameters are affected by different physical
effects.
A clear understanding of the degeneracies that result in particular observables provides physical
insight into the status of light relics with current observations and clarifies how combining
multiple probes breaks these degeneracies.

Beyond the well-established degeneracy between $\omega_r$ and the helium yield $\YHe$, we identified
a partial degeneracy between $\omega_r$, $n_s$, and $\omega_b$, which we coin a ``tilt'' degeneracy
(\cref{sec:tilt_degen}) due to its origin in the scale dependence of the CMB temperature and
polarization spectra.
We also explored how varying the running of the tilt, $\alpha_s$, can mitigate the impact of
small-scale damping induced by additional radiation.
Then we reviewed the fact that, when isolating the background-level effects of extra radiation by fixing
$\omega_r$, the remaining physical effects encoded by the free-streaming fraction $\ffs$ are a phase
and amplitude shift in the acoustic peaks.
As a result, $\ffs$ is strongly anticorrelated with $\theta_s$ and correlated with $A_s$ (as well as
$n_s$ and $\omega_b$, to a lesser extent) via the ``shift'' degeneracy (\cref{sec:shift_degen}).
Neither of the degeneracies with $A_s$ and $\theta_s$ are exact; changes to $\ffs$ only predict a
constant amplitude shift at small scales, and an additive phase shift cannot be exactly reproduced by
scaling $\theta_s$.

Having established the relevant physical effects for the general case featuring arbitrary amounts of fluidlike and free-streaming radiation, we turned to specific new-physics scenarios that introduce new light relics (on top of SM neutrinos) that either are or are not strongly self-interacting.
When interpreted in the $\omega_r$-$\ffs$ parameter space, the two cases differ in whether the
free-streaming fraction increases or decreases with the total radiation density (see
\cref{eqn:perturb_ffs}).
We detailed how additional free-streaming and fluidlike radiation consequently affect the amplitude of the first peak of the CMB primary power spectrum differently.
For models featuring fluidlike radiation, the effects of changing the fraction of pressure-supported
matter and the free-streaming fraction conspire on small scales to hide oscillatory effects: the
former shifts the zero point of the acoustic oscillations, cancelling some of the phase shift
incurred by the latter.
The two effects, however, exacerbate the scale-independent boost in amplitude on small scales but
act in tandem to largely preserve the amplitude of the first acoustic peak.
The first peak thus plays a key role in determining $A_s$ and therefore constraining $\Delta \Nfld$,
as discussed in \cref{sec:f_fs_vs_pressure_supp}.
In contrast, the shift of the acoustic peaks is more severe with additional free-streaming radiation, for which reason constraints on $\Delta \Nfs$ are typically stronger than those on $\Delta \Nfld$.

In \cref{sec:constraints}, we updated constraints on models with additional light relics using the latest \Planck{}, ACT lensing, and DESI data releases, as tabulated in \cref{fig:Ntot-chi-violin,fig:dNfs-dNfld-violin}.
We also studied how constraints degrade upon freeing the primordial helium fraction $\YHe$ or the running of the primordial spectral tilt, $\alpha_s$.
\Planck{} PR4 CMB data alone do not show a preference for additional fluidlike or free-streaming
radiation, with the $\LambdaCDM$ prediction for $\Ntot$ being within about $1 \sigma$ of the posterior
medians across all model extensions and dataset combinations (see \cref{fig:Ntot-chi-violin}).
In general, the data allow for slightly more fluidlike radiation than free-streaming radiation, due
to the aforementioned, differing interplay between the composition of the matter and radiation
content.

Interestingly, we showed in \cref{fig:pr3_vs_pr4_YHe_alpha} that the PR4 reanalysis of the \Planck{}
data prefers a larger value for the helium yield, whose origin \cref{fig:free_f_fs_free_YHe_spag}
attributed to an increase in the polarization amplitude at $\ell \lesssim 1000$ and in the
small-scale damping of temperature anisotropies in PR4 data (both compared to $\LambdaCDM$
predictions and to PR3 data).
Increasing $\YHe$ prolongs recombination, generating a larger polarization signal at last
scattering, and increases damping.
When enforcing consistency with BBN predictions, this preference for a larger helium yield propagates to a preference for greater $\Ntot$ for both fluidlike and free-streaming radiation.
This same feature would likely drive a preference for other $\LambdaCDM$ extensions that modify the
shape of the visibility function, such as a time-varying fine-structure
constant~\cite{Baryakhtar:2024rky}.
However, the data driving this trend are not especially well fit by the model, even with the
additional freedom of $\Ntot$ and $\YHe$: many of the polarization bins in this range skew
multiple standard deviations beyond the support of the posteriors (as projected into the data space),
while the others agree more closely.
The robustness of these features in PR4 data (and any conclusions depending upon them) merit further
scrutiny, especially given their absence in PR3 data.

We also found that recent CMB lensing data from ACT DR6 (combined with \Planck{} PR4) tighten upper limits on models with additional fluidlike radiation, deriving from their preference for a lensing amplitude in excess of $\LambdaCDM$ predictions.
Since lower free-streaming fractions require a smaller amplitude $A_s$ to fit the temperature and polarization data, additional fluidlike radiation is more strongly restricted by the addition of lensing data.
Regardless of the dataset combination or model extension, we find that the fraction of the radiation
density (excluding photons) that is fluidlike is limited to $\Nfld/\Ntot \leq 21\%$ at the $95\%$
level, as seen in \cref{fig:Ntot-chi-violin,fig:dNfs-dNfld-violin}.
This result holds even when $\YHe$ and $\alpha_s$ are free parameters.

Including BAO data from DESI yields a slight preference for extra radiation, deriving from its interplay with the CMB's geometric degeneracy (see \cref{fig:BAO_effects}).
Increasing the radiation density shifts and extends \Planck{}'s preferences along the
geometric degeneracy toward the parameter space preferred by DESI; this phenomenon is more effective
under additional fluidlike radiation, again because its effects that modulate the acoustic peaks
partially cancel.
Using BAO measurements from prior surveys does not yield a similar preference, as DESI's prefer
smaller matter fractions $\Omega_m$ than other observations.
This preference for additional radiation with DESI BAO data lead the CMB to infer larger Hubble
constants $H_0 = 69.8^{+1.2}_{-1.1}~\mathrm{km}/\mathrm{s}/\mathrm{Mpc}$, as observed
in~\cite{Allali:2024cji}.

In \cref{sec:degen_breaking}, we explored how large-scale-structure observations can break both the ``tilt'' degeneracy as well as the degeneracy between $\Ntot$ and $\YHe$.
Full-shape data from spectroscopic galaxy surveys stand to significantly improve inferences of $\Ntot$ and $\YHe$~\cite{Baumann:2017gkg}, a possibility worthy of future investigation (for instance, with new measurements from DESI~\cite{DESI:2024hhd}).
In addition, measurements of the light element abundances constrain the radiation density at higher
temperatures than does the CMB, which can improve measurements on $\Ntot$~\cite{Giovanetti:2021izc,
Giovanetti:2024zce, Giovanetti:2024eff} or probe its possible evolution between nucleosynthesis and
recombination~\cite{Sobotka:2022vrr, Aloni:2023tff, Sobotka:2023bzr}.
The CMB, however, is sensitive to not just the total radiation density but also its
interactions.

In general, the partial degeneracies at play in the CMB can be significantly abated by higher resolution observations from ongoing and future CMB experiments.
Recent data from the Atacama Cosmology Telescope~\cite{ACT:2020gnv} and the South Pole Telescope~\cite{SPTpol:2025kpo}, which observe deeper into the damping tail and offer improved precision in polarization even on moderate scales, should improve meaningfully upon \Planck{}.
Our discussion of fluidlike radiation---namely, the importance of the first acoustic peaks emphasized in \cref{sec:fs_vs_fld}---may shed light on a reported preference for strongly
interacting neutrinos in ACT DR4 data~\cite{Kreisch:2022zxp}: the strongest preferences derive
from results that exclude CMB data other than ACT's high-$\ell$ observations or that include data
from the Wilkinson Microwave Anisotropy Probe~\cite{WMAP:2012fli,WMAP:2012nax} rather than \Planck{}.
Given that DESI's DR1 BAO data are also best accommodated (in combination with \Planck{}) by extra
fluidlike radiation, these scenarios warrant further study with future data.

Finally, we forecasted the CMB-S4 experiments's sensitivity to additional light relic degrees of
freedom (see \cref{fig:s4_vs_pr4_1d}), using its current planned configuration.
These observations will dramatically diminish the degeneracies at play in current data, improving sensitivity to both the existence of light relics as well as the nature of their interactions.
In models where the total radiation and the free-streaming fraction are both allowed to vary, CMB-S4
can reach a sensitivity of $\sigma(\Ntot) \sim 0.03$ and a limit on the fluidlike fraction of new
radiation of $\Nfld / \Ntot \lesssim 4\%$ (at the $95\%$ level).
These future measurements---roughly five times better than present ones---promise an era of
percent-level constraints on not just the existence of new light relics but also their fundamental
nature.

\acknowledgments
We thank Francis-Yan Cyr-Racine, Benjamin Wallisch, Srinivasan Raghunathan, Kimberly Boddy, Cynthia Trendafilova, Nikita A. Zemlevskiy, Jacob W. Crawford, Ella C. Henry, Caio Bastos de Senna Nascimento, Charuhas Shiveshwarker, John Franklin Crenshaw, and Roland C. Farrell for advice and helpful discussions. We also thank Steven Gratton and Erik Rosenberg for providing binned \texttt{CamSpec} spectra.

MMS and ML are supported by the Department of Energy grants DE-SC0023183 and DE-SC0011637. TB was supported by ICSC -- Centro Nazionale di Ricerca in High Performance Computing, Big Data and Quantum Computing, funded by European Union -- NextGenerationEU.
Research at Perimeter Institute is supported in part by the Government of Canada through the Department of Innovation, Science and Economic Development and by the Province of Ontario through the Ministry of Colleges and Universities.
This work was enabled, in part, by the use of advanced computational, storage, and networking infrastructure provided by the Hyak supercomputer system at the University of Washington, which was supported by the UW Student Technology Fee~\cite{uwhyak}. This work made use of the software packages \texttt{corner.py}~\cite{corner}, \texttt{NumPy}~\cite{harris2020array}, \texttt{SciPy}~\cite{2020SciPy-NMeth}, \texttt{matplotlib}~\cite{Hunter:2007}, \texttt{xarray}~\cite{hoyer2017xarray}.

\appendix

\section{Constraining power of different multipole ranges}
\label{sec:diff_multipole_ranges}

\Cref{sec:f_fs_vs_pressure_supp} demonstrates the importance of the height of the first peak plays in constraining $\Nfld$.
To test the impact of this effect on parameter inference, we constrain $\Delta \Nfs$ and $\Delta \Nfld$ using mock CMB temperature and polarization data divided into two multipole ranges: $1000 \leq \ell \leq 2500$, which excludes the first three peaks but captures the damping tail, and $\ell \leq 2500$, which includes the full spectra.
We generate a mock CMB dataset with \Planck{} PR3-like noise and use the \texttt{fake\_planck\_bluebook} likelihood in \montepython{} version 3.6 interfaced with the Boltzmann solver \class{} 3.2.3~\cite{Audren:2012wb,Brinckmann:2018cvx,Blas:2011rf}. The fiducial cosmology takes $\omega_b = 0.02236$, $\omega_c =0.1202$, $100 \theta_s = 1.0409$,
$\ln\left(10^{10}A_s\right) = 3.045$, $n_s=0.965$, $\taureio=0.0544$, and $\Nfs=3.046$ with $\YHe$ fixed to its BBN prediction.
We employ the Metropolis-Hastings algorithm in \montepython{} for sampling and a Gelman-Rubin convergence criterion of $R-1 \lesssim 0.01$, leading to effective sample sizes of at least 20,000.

\begin{figure}
    \includegraphics[width=0.5\textwidth]{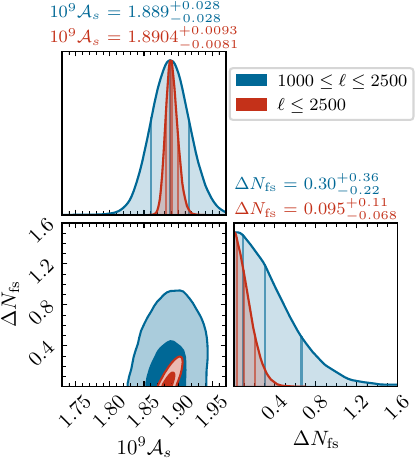}
    \includegraphics[width=0.5\textwidth]{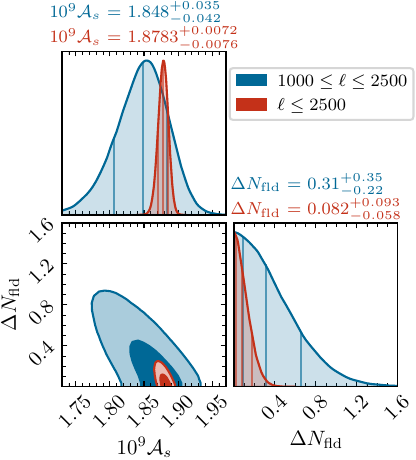}
    \caption{
        Posterior distribution for $\Astau$ and additional radiation ($\Delta \Nfs$, left, and $\Delta \Nfld$, right) for a mock-\Planck{} likelihood over $\ell \leq 2500$ (red) and restricted to $1000 \leq \ell \leq 2500$ (blue).
        Results are presented as in \cref{fig:pr3_vs_pr4_YHe_alpha}.
    }
    \label{fig:A_s_tau_diff_ranges}
\end{figure}

In \cref{fig:A_s_tau_diff_ranges}, we plot the posterior distributions for $\Delta \Nfs$ and $\Delta \Nfld$ (each varied independently) along with $\mathcal{A}_s = \Astau$, since this combination of parameters controls the overall amplitude of the spectra.
With additional free-streaming radiation, posteriors for $1000\leq\ell\leq2500$ and for the full multipole range are centered on essentially the same value of $\mathcal{A}_s$, as they show little degeneracy between $\mathcal{A}_s$ and $\Delta \Nfs$.
Contrast this result with the case with additional fluidlike radiation, where the medians are
substantially offset from each other due to a clear degeneracy within high-$\ell$ data; the
posterior over $\mathcal{A}_s$ for the high-$\ell$ subset is also substantially broader under
additional fluidlike radiation.

Evidently, the first peaks play much more of a role in breaking degeneracies with $\mathcal{A}_s$ in
models with additional fluidlike radiation, as anticipated from \cref{sec:f_fs_vs_pressure_supp}
based on the differing interplay between changes to the pressure-supported matter fraction and the
free-streaming fraction.
Notably, the posteriors over $\Delta \Nx$ are extremely similar when using only high-$\ell$ data, while upper limits from the full multipole range are slightly tighter on $\Delta \Nfld$---namely,
$\Delta \Nfld < 0.25$ and $\Delta \Nfs < 0.29$ at the $95\%$ level.
This comparison suggests that the impacts of the free-streaming fraction (\cref{sec:vary-ffs}) are
only relevant insofar as they mediate the spectra at low and high multipole.

\section{Supplementary results}
\label{app:supplementary-results}

\begin{figure}[t!]
    \centering
    \includegraphics[width=\textwidth]{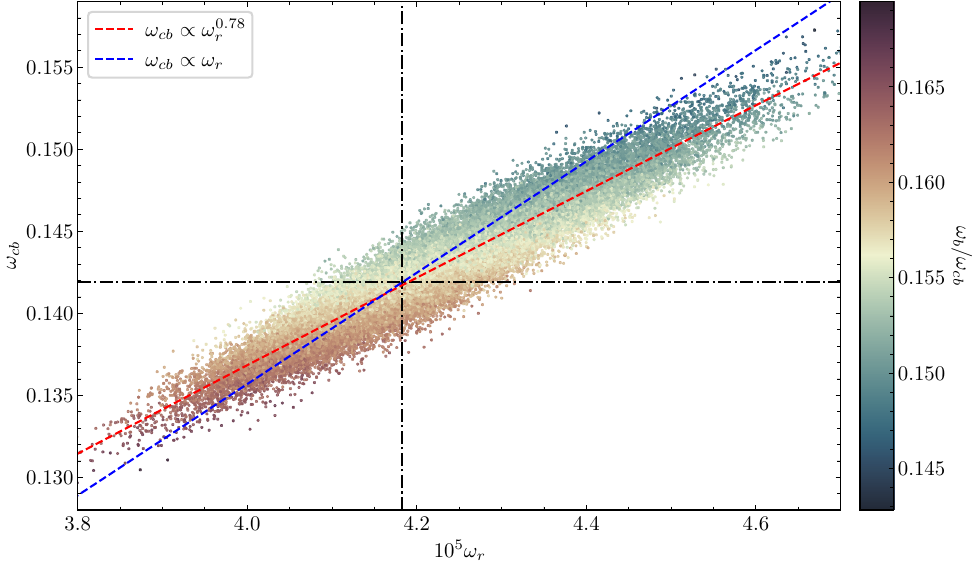}
    \caption{
    Scatter plot of a posterior varying $\Ntot$ and $\ffs$ in the $\omega_r$-$\omega_{cb}$ plane, colored by the corresponding value of $\omega_b/\omega_{cb}$.
    Results set $\YHe$ to its BBN prediction and use the \Planck{} PR4 likelihoods with no lensing data.
    The blue dashed line indicates the linear correspondence that fixes the redshift of matter-radiation equality, while the red dashed line displays a numerical fit, demonstrating the sublinear relationship.
    The vertical dot-dashed line indicates the value for $\omega_r$ with the neutrino density predicted by the Standard Model.
    The horizontal dot-dashed line marks the best-fit $\LambdaCDM$ value of $\omega_{cb}$ derived from PR4 data.
    }
    \label{fig:pressure_supp_mcmc}
\end{figure}

\begin{figure}[t!]
    \centering
    \includegraphics[width=\textwidth]{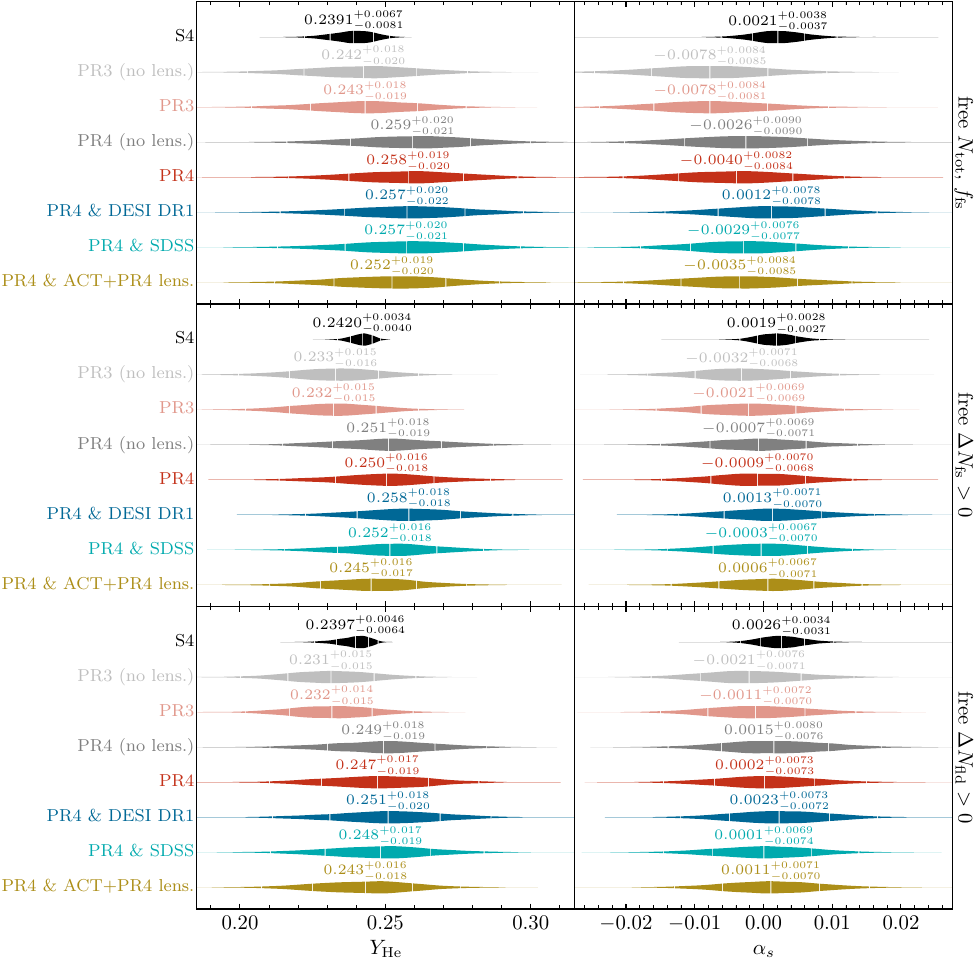}
    \caption{
        Marginal posterior distributions over $\YHe$ and $\alpha_s$ in models where they freely vary. The S4 posteriors correspond to forecasts assuming the fiducial model detailed in
        \cref{sec:CMB-S4} (see \cref{fig:s4_vs_pr4_1d}), while all others derive from current data
        (see \cref{fig:Ntot-chi-violin,fig:dNfs-dNfld-violin}).
        Note that the DESI and SDSS dataset combinations include PR4 lensing data.
        Results are presented as in \cref{fig:Ntot-chi-violin}.
    }
    \label{fig:YHe-alpha_s-violin-all}
\end{figure}

\Cref{fig:pressure_supp_mcmc} shows that posteriors from \Planck{} data exhibit a degeneracy
$\omega_r \approx \omega_{cb}^{0.78}$ driven by a compromise between fixing the scale factor of matter-radiation equality $\aeq$ and the pressure-supported matter fraction $\omega_b/\omega_{cb}$.
We also provide constraints on $\LambdaCDM$, models with additional free-streaming or fluidlike radiation, and models where $\Ntot$ and $\ffs$ are allowed to vary.
\Cref{fig:YHe-alpha_s-violin-all} depicts the marginalized posteriors over $\YHe$ and $\alpha_s$ for models where they vary for all dataset combinations considered in this work.
\Cref{tab:TTTEEE_constraints,tab:+lensing_constraints,tab:+act_constraints,tab:+desi_constraints}
provide constraints on the $\LambdaCDM$ parameters, $\Ntot$, $\ffs$, and the fraction of radiation (outside
the photons) that is fluidlike, $\Nfld/\Ntot$, as well as the derived parameters $H_0$, $\rdrag$, $\Omega_m$,
and the amplitude of matter fluctuations, $\sigma_8$.
\Cref{tab:TTTEEE_constraints} reports results using only the \Planck{} PR4 temperature and polarization data, while \cref{tab:+lensing_constraints} includes PR4 lensing data, \cref{tab:+act_constraints} also includes ACT DR6 lensing data, and \cref{tab:+desi_constraints} also includes DESI DR1 BAO data.
Note that $\YHe$ is fixed to the BBN prediction in these results.
\begin{table}[t]
    \renewcommand{\arraystretch}{1.5}
    \centering
    \begin{tabular}[t]{lcccc}

        \toprule
         &$\LambdaCDM$ & $\Delta \Nfs$ & $\Delta \Nfld$ & $\Ntot, \ffs$
        \\
        \midrule
        $100\,\theta_s$ & $1.04175 ^{+0.00025}_{-0.00024}$ & $1.04144 ^{+0.00031}_{-0.00035}$ & $1.04201 ^{+0.00032}_{-0.0003}$ & $1.04251 ^{+0.00072}_{-0.00066}$
        \\
        $\omega_b$ & $0.02217 ^{+0.00014}_{-0.00014}$ & $0.02227 ^{+0.00016}_{-0.00015}$ & $0.02239 ^{+0.00022}_{-0.00019}$ & $0.02233 ^{+0.00024}_{-0.00023}$
        \\
        $\omega_c$ & $0.1196 ^{+0.0012}_{-0.0012}$ & $0.1216 ^{+0.0022}_{-0.0018}$ & $0.1226 ^{+0.0028}_{-0.0022}$ & $0.1207 ^{+0.0037}_{-0.0036}$
        \\
        $\ln\left(10^{10} A_s\right)$ & $3.046 ^{+0.013}_{-0.013}$ & $3.052 ^{+0.014}_{-0.014}$ & $3.041 ^{+0.014}_{-0.014}$ & $3.031 ^{+0.017}_{-0.018}$
        \\
        $n_s$ & $0.963 ^{+0.0043}_{-0.0042}$ & $0.9679 ^{+0.006}_{-0.0052}$ & $0.966 ^{+0.0047}_{-0.0046}$ & $0.9602 ^{+0.0082}_{-0.0081}$
        \\
        $\tau_{\mathrm{reio}}$ & $0.0573 ^{+0.0062}_{-0.006}$ & $0.0578 ^{+0.0063}_{-0.0061}$ & $0.0581 ^{+0.0063}_{-0.0061}$ & $0.0579 ^{+0.0063}_{-0.0061}$
        \\
        $ \Delta \Ntot $& --- & $< 0.39$ & $< 0.55$ & $0.09 ^{+0.25}_{-0.23}$
        \\
        $f_{\mathrm{fs}}$ &  --- &  --- &  --- & $0.378 ^{+0.024}_{-0.026}$
        \\
        $\Nfld/\Ntot$  & --- & --- & $< 0.15$ & $ < 0.20$
        \\
        \midrule
        $H_0$ [km/s/Mpc] & $67.26 ^{+0.55}_{-0.53}$ & $68.2 ^{+1.1}_{-0.8}$ & $69.0 ^{+1.7}_{-1.2}$ & $68.3 ^{+1.9}_{-1.7}$
        \\
        $\rdrag$ [Mpc] & $147.43 ^{+0.28}_{-0.27}$ & $146.2 ^{+0.9}_{-1.4}$ & $145.5 ^{+1.3}_{-1.8}$ & $146.5 ^{+2.3}_{-2.3}$
        \\
        $\Omega_m$ & $0.3148 ^{+0.0074}_{-0.0076}$ & $0.3104 ^{+0.0077}_{-0.0079}$ & $0.305 ^{+0.009}_{-0.01}$ & $0.308 ^{+0.011}_{-0.011}$
        \\
        $\sigma_8$ & $0.8102 ^{+0.0064}_{-0.0064}$ & $0.8158 ^{+0.008}_{-0.0073}$ & $0.8121 ^{+0.0066}_{-0.0065}$ & $0.805 ^{+0.011}_{-0.011}$
        \\
        \bottomrule
    \end{tabular}
    \caption{Constraints on the extensions to $\LambdaCDM$ considered in this work using the \Planck{} PR4 TT, TE, and EE likelihoods (with no lensing included), with $\YHe$ fixed to the BBN prediction. We present the median and $\pm 1 \sigma$ quantiles for most parameters and $95\%$ upper limits for $\Nfld/\Ntot$ and $\Delta \Nx$. In all cases, $\Delta \Ntot = \Ntot - 3.044$ (whether from free-streaming radiation, fluidlike radiation, or both).}
    \label{tab:TTTEEE_constraints}
\end{table}

\begin{table}[t]
    \renewcommand{\arraystretch}{1.5}
    \centering
    \begin{tabular}[t]{lcccc}
        \toprule
         &$\LambdaCDM$ & $\Delta \Nfs$ & $\Delta \Nfld$ & $\Ntot, \ffs$
        \\
        \midrule
        $100\,\theta_s$ & $1.04175 ^{+0.00024}_{-0.00024}$ & $1.04146 ^{+0.00031}_{-0.00035}$ & $1.04196 ^{+0.0003}_{-0.00029}$ & $1.04243 ^{+0.00066}_{-0.00061}$
        \\
        $\omega_b$ & $0.02217 ^{+0.00014}_{-0.00014}$ & $0.02227 ^{+0.00015}_{-0.00015}$ & $0.02235 ^{+0.0002}_{-0.00017}$ & $0.02228 ^{+0.00022}_{-0.00021}$
        \\
        $\omega_c$ & $0.1196 ^{+0.001}_{-0.001}$ & $0.1215 ^{+0.0021}_{-0.0017}$ & $0.1223 ^{+0.0026}_{-0.002}$ & $0.1206 ^{+0.0034}_{-0.0035}$
        \\
        $\ln\left(10^{10} A_s\right)$ & $3.047 ^{+0.012}_{-0.012}$ & $3.053 ^{+0.013}_{-0.013}$ & $3.044 ^{+0.012}_{-0.012}$ & $3.035 ^{+0.016}_{-0.016}$
        \\
        $n_s$ & $0.963 ^{+0.004}_{-0.0039}$ & $0.9675 ^{+0.0057}_{-0.005}$ & $0.965 ^{+0.0042}_{-0.0042}$ & $0.9592 ^{+0.008}_{-0.0079}$
        \\
        $\tau_{\mathrm{reio}}$ & $0.0577 ^{+0.0062}_{-0.0059}$ & $0.0582 ^{+0.006}_{-0.0059}$ & $0.0589 ^{+0.0062}_{-0.0059}$ & $0.0586 ^{+0.0062}_{-0.006}$
        \\
        $\Delta \Ntot $ & --- & $<0.37$ & $<0.47$ & $0.06 ^{+0.23}_{-0.22}$
        \\
        $f_{\mathrm{fs}}$ & --- & --- & --- & $0.381 ^{+0.022}_{-0.025}$
        \\
        $\Nfld/\Ntot$  & --- & --- & $<0.13$ & $<0.18$
        \\
        \midrule
        $H_0$ [km/s/Mpc] & $67.26 ^{+0.48}_{-0.46}$ & $68.13 ^{+1.0}_{-0.76}$ & $68.6 ^{+1.4}_{-1.0}$ & $67.9 ^{+1.7}_{-1.6}$
        \\
        $\rdrag$ [Mpc] & $147.43 ^{+0.24}_{-0.24}$ & $146.2 ^{+0.8}_{-1.3}$ & $145.7 ^{+1.2}_{-1.6}$ & $146.8 ^{+2.2}_{-2.2}$
        \\
        $\Omega_m$ & $0.3149 ^{+0.0064}_{-0.0064}$ & $0.3108 ^{+0.0071}_{-0.007}$ & $0.3079 ^{+0.0078}_{-0.0083}$ & $0.311 ^{+0.0094}_{-0.0092}$
        \\
        $\sigma_8$ & $0.8107 ^{+0.0051}_{-0.0052}$ & $0.816 ^{+0.0071}_{-0.0063}$ & $0.8136 ^{+0.0057}_{-0.0055}$ & $0.8067 ^{+0.0096}_{-0.01}$
        \\
        \bottomrule
    \end{tabular}
    \caption{Constraints on the extensions to $\LambdaCDM$ considered in this work using the \Planck{} PR4 TT, TE, EE, and lensing likelihoods. Results are presented as in \cref{tab:TTTEEE_constraints}.}
    \label{tab:+lensing_constraints}
\end{table}

\begin{table}[t]
    \renewcommand{\arraystretch}{1.5}
    \centering
    \begin{tabular}[t]{lcccc}

        \toprule
         &$\LambdaCDM$ & $\Delta \Nfs$ & $\Delta \Nfld$ & $\Ntot, \ffs$
        \\
        \midrule
        $100\,\theta_s$ & $1.04173 ^{+0.00024}_{-0.00024}$ & $1.04146 ^{+0.0003}_{-0.00033}$ & $1.04189 ^{+0.00028}_{-0.00027}$ & $1.04235 ^{+0.00061}_{-0.00057}$
        \\
        $\omega_b$ & $0.02217 ^{+0.00014}_{-0.00014}$ & $0.02226 ^{+0.00015}_{-0.00015}$ & $0.02231 ^{+0.00018}_{-0.00016}$ & $0.02222 ^{+0.00021}_{-0.0002}$
        \\
        $\omega_c$ & $0.1198 ^{+0.0011}_{-0.0011}$ & $0.1216 ^{+0.002}_{-0.0016}$ & $0.1218 ^{+0.0022}_{-0.0017}$ & $0.1199 ^{+0.0032}_{-0.0032}$
        \\
        $\ln\left(10^{10} A_s\right)$ & $3.05 ^{+0.011}_{-0.011}$ & $3.056 ^{+0.012}_{-0.012}$ & $3.05 ^{+0.011}_{-0.011}$ & $3.041 ^{+0.015}_{-0.015}$
        \\
        $n_s$ & $0.9631 ^{+0.004}_{-0.004}$ & $0.9673 ^{+0.0056}_{-0.005}$ & $0.9649 ^{+0.0043}_{-0.0042}$ & $0.9587 ^{+0.0079}_{-0.0077}$
        \\
        $\tau_{\mathrm{reio}}$ & $0.0584 ^{+0.0061}_{-0.0059}$ & $0.059 ^{+0.0061}_{-0.0059}$ & $0.0597 ^{+0.0063}_{-0.006}$ & $0.0593 ^{+0.0062}_{-0.006}$
        \\
        $\Delta \Ntot$ & --- & $<0.34$ & $<0.37$ & $0.00 ^{+0.21}_{-0.21}$
        \\
        $f_{\mathrm{fs}}$ & --- & --- & --- & $0.383 ^{+0.02}_{-0.022}$
        \\

        $\Nfld/\Ntot$  & --- & --- & $< 0.11$ & $ < 0.15$
        \\
        \midrule
        $H_0$ [km/s/Mpc] & $67.18 ^{+0.49}_{-0.48}$ & $67.98 ^{+0.96}_{-0.73}$ & $68.2 ^{+1.2}_{-0.8}$ & $67.4 ^{+1.6}_{-1.5}$
        \\
        $\rdrag$ [Mpc] & $147.38 ^{+0.24}_{-0.25}$ & $146.3 ^{+0.8}_{-1.2}$ & $146.1 ^{+0.9}_{-1.4}$ & $147.3 ^{+2.1}_{-2.0}$
        \\
        $\Omega_m$ & $0.316 ^{+0.0067}_{-0.0067}$ & $0.3123 ^{+0.007}_{-0.0072}$ & $0.3108 ^{+0.0074}_{-0.0077}$ & $0.3146 ^{+0.0092}_{-0.0091}$
        \\
        $\sigma_8$ & $0.8128 ^{+0.0046}_{-0.0045}$ & $0.8177 ^{+0.0066}_{-0.0057}$ & $0.8156 ^{+0.0052}_{-0.005}$ & $0.8086 ^{+0.0093}_{-0.0094}$
        \\
        \bottomrule
    \end{tabular}
    \caption{Constraints on the extensions to $\LambdaCDM$ considered in this work using the \Planck{} PR4 TT, TE, and EE likelihoods along with the \Planck{} PR4 and ACT DR6 lensing likelihoods. Results are presented as in \cref{tab:TTTEEE_constraints}.}
    \label{tab:+act_constraints}
\end{table}

\begin{table}[t]
    \renewcommand{\arraystretch}{1.5}
    \centering
    \begin{tabular}[t]{lcccc}
        \toprule
         &$\LambdaCDM$ & $\Delta \Nfs$ & $\Delta \Nfld$ & $\Ntot, \ffs$
        \\
        \midrule
        $100\,\theta_s$ & $1.04188 ^{+0.00023}_{-0.00023}$ & $1.04141 ^{+0.00036}_{-0.0004}$ & $1.04213 ^{+0.00028}_{-0.00027}$ & $1.04247 ^{+0.00071}_{-0.00067}$
        \\
        $\omega_b$ & $0.02228 ^{+0.00013}_{-0.00013}$ & $0.02239 ^{+0.00015}_{-0.00015}$ & $0.02249 ^{+0.00018}_{-0.00017}$ & $0.02247 ^{+0.00018}_{-0.00018}$
        \\
        $\omega_c$ & $0.1182 ^{+0.00081}_{-0.00081}$ & $0.1215 ^{+0.0026}_{-0.0022}$ & $0.1232 ^{+0.0028}_{-0.0028}$ & $0.1224 ^{+0.0033}_{-0.0035}$
        \\
        $\ln\left(10^{10} A_s\right)$ & $3.052 ^{+0.012}_{-0.012}$ & $3.059 ^{+0.013}_{-0.013}$ & $3.045 ^{+0.013}_{-0.013}$ & $3.04 ^{+0.017}_{-0.017}$
        \\
        $n_s$ & $0.9664 ^{+0.0036}_{-0.0036}$ & $0.9722 ^{+0.0057}_{-0.0052}$ & $0.9675 ^{+0.0038}_{-0.0038}$ & $0.9643 ^{+0.0072}_{-0.0072}$
        \\
        $\tau_{\mathrm{reio}}$ & $0.0615 ^{+0.006}_{-0.006}$ & $0.0611 ^{+0.006}_{-0.0058}$ & $0.0613 ^{+0.0062}_{-0.0059}$ & $0.0614 ^{+0.0061}_{-0.006}$
        \\
        $\Delta \Ntot$ & --- & $<0.48$ & $<0.56$ & $0.23 ^{+0.20}_{-0.20}$
        \\
        $f_{\mathrm{fs}}$ & --- & --- & --- & $0.381 ^{+0.025}_{-0.027}$
        \\
        $\Nfld/\Ntot$ & ---&  ---& $<0.15$ & $<0.21$
        \\
        \midrule
        $H_0$ [km/s/Mpc] & $67.9 ^{+0.37}_{-0.37}$ & $69.1 ^{+1.1}_{-0.8}$ & $69.8 ^{+1.2}_{-1.1}$ & $69.5 ^{+1.3}_{-1.3}$
        \\
        $\rdrag$ [Mpc] & $147.69 ^{+0.21}_{-0.21}$ & $145.7 ^{+1.3}_{-1.6}$ & $144.8 ^{+1.6}_{-1.6}$ & $145.2 ^{+2.0}_{-1.9}$
        \\
        $\Omega_m$ & $0.3061 ^{+0.0049}_{-0.0049}$ & $0.3025 ^{+0.0054}_{-0.0055}$ & $0.3 ^{+0.0057}_{-0.0057}$ & $0.3009 ^{+0.006}_{-0.006}$
        \\
        $\sigma_8$ & $0.8088 ^{+0.0052}_{-0.005}$ & $0.8173 ^{+0.0083}_{-0.0074}$ & $0.814 ^{+0.0061}_{-0.0059}$ & $0.8101 ^{+0.0096}_{-0.0097}$
        \\

        \bottomrule
    \end{tabular}
    \caption{Constraints on the extensions to $\LambdaCDM$ considered in this work using the \Planck{} PR4 TT, TE, EE, and lensing likelihoods as well as the DESI BAO data. Results are presented as in \cref{tab:TTTEEE_constraints}.}
    \label{tab:+desi_constraints}
\end{table}

\bibliographystyle{JHEP}
\bibliography{refs}
\end{document}